\documentclass[mksc]{informs4}

\usepackage{eqndefns-left}%
\RequirePackage{tgtermes}
\RequirePackage{newtxtext}
\RequirePackage{newtxmath}
\RequirePackage{bm}
\RequirePackage{endnotes}

\OneAndAHalfSpacedXII%

\usepackage{natbib}
\bibpunct[, ]{(}{)}{,}{a}{}{,}%
\def\bibfont{\small}%

\EquationsNumberedThrough
\TheoremsNumberedThrough
\ECRepeatTheorems

\MANUSCRIPTNO{}

\usepackage[utf8]{inputenc}
\usepackage[T1]{fontenc}
\usepackage{hyperref}
\usepackage{url}
\hypersetup{hypertexnames=false}
\usepackage{booktabs}
\usepackage{longtable}
\usepackage{multirow}
\usepackage{amsfonts}
\usepackage{amsmath}
\usepackage{nicefrac}
\usepackage{microtype}
\usepackage{graphicx}
\usepackage{xcolor}
\usepackage{float}
\usepackage{tikz}
\usepackage{xcolor}
\usetikzlibrary{positioning,fit,backgrounds,decorations.pathreplacing,calc}

\newcommand{\safeincludegraphics}[2][]{%
  \IfFileExists{#2}{\includegraphics[#1]{#2}}{%
    \IfFileExists{figures/#2}{\includegraphics[#1]{#2}}{%
      \fbox{\begin{minipage}[c][1.8in][c]{0.90\linewidth}\centering\small Missing figure file: \texttt{\detokenize{#2}}\end{minipage}}%
    }%
  }%
}

\newenvironment{promptblock}
  {\begin{quote}\small\upshape``\ignorespaces}
  {\unskip''\end{quote}}
\usepackage{placeins}
\usepackage{array}
\usepackage[htt]{hyphenat}%
\providecommand{\singlespacing}{\def\baselinestretch{1}\selectfont}%

\graphicspath{{figures/}}

\begin{document}

\RUNAUTHOR{Deng, Liu, Toubia, and Jain}
\RUNTITLE{AI-Moderated Interviews for Market Research}

\TITLE{AI-Moderated Interviews for Market Research and Digital Twins Calibration}

\ARTICLEAUTHORS{%
\AUTHOR{}
\AFF{}
}

\renewcommand{\theARTICLEAUTHORS}{%
\vspace{10pt}
\begin{center}
{\fontsize{13}{15}\selectfont\rmfamily\bfseries
Yuting Deng,\textsuperscript{1}\quad Jingxuan Liu,\textsuperscript{1}\quad Olivier Toubia,\textsuperscript{1}\quad Naman Jain\textsuperscript{2}}\\[8pt]
{\normalsize \textsuperscript{1}Columbia Business School, Columbia University}\\[3pt]
{\normalsize \textsuperscript{2}The Wharton School, University of Pennsylvania}\\[6pt]
{\small \texttt{\{yd2721, jliu27, ot2107\}@gsb.columbia.edu}\quad\texttt{namanr999@gmail.com}}
\end{center}
\vspace{4pt}
}

\ABSTRACT{
AI-moderated interviews are emerging as a scalable market-research method for generating consumer insights and building consumer ``digital twins.'' Yet it remains unclear whether they match human-moderated interviews and whether they improve on simpler, static data collection methods. In a pre-registered study with three industry partners, the authors randomly assign participants to AI-moderated interviews ($N=139$), static interviews ($N=154$), or interviews conducted by professional human moderators ($N=24$). AI moderation matches human moderation in depth, covers more themes, and, at equal budget, recovers more customer needs than either alternative, although participants sound more emotionally engaged with a human interviewer. The authors then build a digital twin of each participant and evaluate it against the participant’s own held-out responses to six real-world marketing stimuli. Twins built from AI-moderated interviews predict consumer responses better than demographics-only personas, but no better than twins built from static interviews: the richer conversation does not necessarily yield more accurate predictions. Comparing the open-ended explanations provided by participants and their twins shows that prediction errors arise when twins reason differently than the people they represent and when questions fall far outside what the interview covered. AI moderation thus pays off most for discovering customer needs, and less for predicting customer responses.
}

\KEYWORDS{AI-moderated interviews; qualitative market research; digital twins; large language models; voice of the customer}

\maketitle

\section{Introduction}
\label{sec:intro}

Customer research is central to firms' decisions about product positioning, advertising, and customer experience, but it faces a persistent tradeoff. Survey-based methods are fast, standardized, and scalable, yet they often miss the ``why'' behind consumer choices \citep{anugraha2026sparkme,korst2026howaihelps}. Human-moderated interviews can uncover richer narratives and more diagnostic insights, but they are slow, costly, and difficult to scale. Recent advances in generative AI promise to relax this tradeoff: AI-powered interviewers can conduct adaptive, open-ended conversations with consumers at a much lower marginal cost, making it possible to bring qualitative insight earlier and more frequently into marketing decisions \citep{chopra2026qualitative,geiecke2024conversations,korst2026howaihelps,wuttke2025aiconversational}. This promise has generated substantial attention and investment: AI-native qualitative research startups such as Outset, Listen Labs, and Simile have raised hundreds of millions of dollars in venture funding  \citep{korst2026howaihelps}. Major LLM companies are also beginning to use AI-moderated interviews at scale; for example, Anthropic has launched large-scale AI-moderated interview programs \citep{anthropic2025interviewer}.

Despite growing interest and practitioner adoption, academic evidence on the performance of AI-moderated interviews in market research remains limited. Two broad questions are especially important. First, how do AI-moderated interviews perform as a market-research method relative to established alternatives such as professional human-moderated interviews and static open-ended interviews? Second, what value does the resulting qualitative data provide for downstream marketing applications such as consumer digital twins? We address these questions in collaboration with an AI-moderated interview platform, a professional market research firm, and a B2C company, allowing us to evaluate AI moderation in a real-world market research setting. We first compare the interview approaches on the quality of the data they generate and their efficiency in uncovering customer needs. We then examine whether the resulting interview data from AI-moderated versus static interviews improve digital twins' ability to predict consumers' responses to novel marketing stimuli. Together, this work aims to clarify the role AI moderation can play relative to traditional approaches in qualitative market research and whether the additional richness of adaptive interviewing translates into greater downstream predictive value.

\subsection{AI-Moderated Interviews for Market Research}

AI-moderated interviews sit between traditional surveys and human-moderated qualitative interviews. Compared with static surveys, qualitative interviews can adaptively ask clarifying and follow-up questions based on what consumers say \citep{arsel2017asking}. Adaptive surveys such as adaptive conjoint analysis also tailor questions to respondents, but they estimate pre-specified utility functions from structured data \citep{johnson1987aca,toubia2003fastpolyhedral,toubia2004polyhedral}. Qualitative interviews instead require moderators to interpret open-ended responses and decide what to ask next. However, human moderators are costly and can charge \$200--\$400/hour (based on communications with GBK Collective, one of our industry partners). 

Compared with human interviews, AI-moderated interviews promise lower cost per interview (approximately 20\% of a human-moderated interview, based on market data from our partners), greater scalability, and consistent implementation of a topic guide \citep{anugraha2026sparkme}. However, despite its promised benefits, AI moderation may introduce new threats to validity. For example, it remains unclear to what extent respondents are comfortable talking to LLMs \citep{defreitas2024aicompanions,kim2022personalquestion,luo2019machines}. LLM hallucinations may also cause AI moderators to ask inappropriate follow-up questions or introduce fabricated content, potentially degrading data quality and requiring validation and human oversight \citep{blanchard2025newtools,schroeder2025llmqualitative}. In addition, LLM-based moderators can add response latency to the conversation \citep{wuttke2025aiconversational}. 

Recent evidence, coming mostly from fields adjacent to marketing, is nonetheless encouraging. Across economic, political, and occupational settings, AI-led interviews receive high quality ratings \citep{geiecke2024conversations, anugraha2026sparkme}, elicit richer responses than open-ended survey questions \citep{geiecke2024conversations, chopra2026qualitative}, and uncover mental models and themes that a fixed set of open-ended questions misses \citep{chopra2026qualitative}. Comparisons with human interviewers have also been favorable: sociologists rate voice-based AI interviews close to face-to-face interviews by trained human interviewers \citep{geiecke2024conversations}. A small study ($N=11$ interviews) finds comparable response quality, but participants rated the AI interview as less interesting and less satisfying \citep{wuttke2025aiconversational}. And in a large field experiment on job interviews, voice AI agents outperform professional recruiters on downstream hiring outcomes \citep{jabarian2026voice}. In marketing, \citet{arora2025hybrids} show that LLMs can moderate in-depth interviews with synthetic respondents and benchmark the resulting data against a human-moderated study of real consumers. 

Despite this encouraging initial evidence, at least two questions remain on the use of AI-moderated interviews for market research. First, it is unclear how AI moderation compares with professional human moderators when both interview real consumers about a real category and brand. \citet{arora2025hybrids}, for example, compare AI-moderated interviews of synthetic personas to human-moderated interviews of human consumers. We complement their work by deploying an AI moderator with real consumers, so that respondent type is held constant and only the moderator varies. Second, outside of hiring \citep{jabarian2026voice}, existing studies evaluate interviews mainly through process measures, such as topic coverage, thematic richness, and participant ratings. For market researchers, however, an interview is valuable to the extent that it efficiently yields decision-relevant insight, such as customer needs \citep{griffin1993}.%

Accordingly, our first goal is to evaluate AI moderation as a market-research method by benchmarking it against two alternatives. We compare AI-moderated interviews with interviews conducted by professional human moderators, to assess how closely AI reproduces the substantive and experiential qualities of traditional qualitative research. We also compare AI-moderated interviews with static interviews, which pose fixed open-ended questions with the same objectives on the same platform and in the same audio modality, to isolate the value of adaptive probing. All conditions cover topics typical of consumer research, including purchase journeys, brand perceptions, and prior brand interactions. We compare the approaches on interview volume, thematic breadth and depth, participants' speech-based affect and prosody, and the customer needs uncovered, both per interview and under a fixed research budget.

\subsection{AI-Moderated Interviews as Training Data for Digital Twins}

One of the major downstream use cases of AI-moderated interviews in marketing is the creation of digital twins: LLM-based representations of specific individuals, built from their own data, such as qualitative interviews or extensive survey responses \citep{park2024agents1000, toubia2025twin2k500}. The twin is then used to predict how that individual would respond to new questions not included in these data. Digital twins have attracted substantial attention as a potential tool for consumer research because they promise to let firms simulate consumer responses at greater scale and lower cost than repeatedly collecting new data from human respondents. Companies such as Outset and Listen Labs offer interview-grounded digital twins, while Aaru, Simile, and Expected Parrot similarly provide AI-based simulations of human responses to surveys, products, pricing, marketing messages, and hypothetical scenarios \citep{outset2026twins,listenlabs2026twins,aaru2026simulation,simile2026simulation,expectedparrot2026simulation}.

However, whether these simulations can accurately predict human responses remains an important empirical question. Existing evidence suggests that adding more person-specific information does not necessarily translate to more accurate digital twin predictions. For example, across 19 pre-registered studies, \citet{peng2026funhouse} find that digital twins built on extensive survey data (combined with demographics) are not more accurate than simpler personas trained on demographics data alone. This raises the question of whether richer forms of consumer data can provide information that structured survey responses and demographic information fail to capture.

Qualitative interview data may offer one such source of information to enrich digital twins. \citet{park2024agents1000} conduct two-hour AI-moderated interviews with more than 1,000 U.S. adults and build an LLM agent for each participant from their transcript. These agents predict participants' held-out responses to the General Social Survey, personality inventories, economic games, and behavioral experiments more accurately than agents built from demographics alone, suggesting that open-ended self-reports contain information beyond demographics that improves downstream predictions. However, at least two aspects of  \citeauthor{park2024agents1000}'s (\citeyear{park2024agents1000}) study call for additional research. First, the interview protocol covers many of the same life domains as the validation measures, such as political views, values, and family background. As a result, some of the predictive gain may reflect leakage, whereby LLMs are evaluated on outcomes that are closely related to information contained in their training data \citep{ludwig2026large}. Second, their interviews follow a long, largely scripted protocol, and the study has no non-adaptive comparison condition. It therefore cannot isolate whether adaptive probing itself contributes to predictive accuracy.

More broadly, prior research on individual-level digital twins has largely focused on social-science domains, examining whether they can reproduce individual traits \citep{toubia2025twin2k500}, attitudes \citep{park2024agents1000}, and behavior on canonical experimental tasks \citep{peng2026funhouse}. Little is known about how digital twins perform in real-world marketing settings, where firms may seek to predict consumers’ responses to novel marketplace stimuli such as advertisements and claims. In particular, it remains unclear what kinds of qualitative consumer data are most useful for constructing digital twins and whether richer, more adaptive interview data translate into more accurate predictions of managerially relevant outcomes.

Hence, the second goal of this paper is to examine whether qualitative data collected through AI-moderated interviews improve digital twins' predictions of managerially relevant outcomes. We construct twins from each participant's own interview about the product category and brand, and evaluate them against that participant's held-out responses to new tasks: real-world advertisements, mailers, and marketing claims provided by one of our three industry partners. Because the stimuli were not discussed in the interviews, this design not only alleviates the leakage concern above, but also tests how far interview-based twins can extrapolate. We compare twins built from AI-moderated interviews, static interviews, and demographics alone, which allows us to test whether the richness that AI moderation adds to interviews also translates into more accurate predictions. We further compare open-ended explanations from participants and their twins to diagnose where predictions fail.

\subsection{The Current Research}

We address our two goals in a pre-registered, between-subjects study of consumers (N=317) in the windows-and-doors home-improvement category (AsPredicted: \url{https://aspredicted.org/wb9sj2.pdf}, \url{https://aspredicted.org/3b3293.pdf}). Participants were randomly assigned to an AI-moderated or a static interview; a third group, recruited separately from the same online pool, completed the same interview guide with professional human moderators.\footnote{Our initial pre-registration only included the AI-moderated and static conditions. The human-moderated condition was added later based on peer feedback, and pre-registered separately.} To ground our investigation in real-world marketing practice, we collaborated with three industry partners. A major North American building-products company provided a relevant market research context and real marketing stimuli. An AI startup specializing in AI-moderated interviews provided the AI platform for conducting state-of-the-art AI-moderated interviews (the same platform was also used for static interviews). GBK Collective, a marketing strategy and insights consultancy with deep expertise in qualitative and quantitative research, served as our qualitative research partner and conducted the human-moderated interviews. All conditions covered the same topics about the product category and the partner brand. Because human-moderated interviews are substantially more expensive and difficult to manage, we matched conditions on market cost (about \$5{,}000 per condition) rather than sample size. 

We find that AI-moderated interviews elicit longer, broader, and deeper responses than static interviews without reducing participant enjoyment or comfort. Compared with professional moderators, AI moderation achieves comparable depth and broader thematic coverage, and under a fixed budget it recovers the most customer needs. However, human-moderated interviews retain an engagement advantage: participants sound more positive, activated, and assertive when speaking live to a human vs. an AI moderator. We detect no differences between AI and static interviews in the measured speech-based affect. Thus, AI moderation matches or exceeds human moderation on interview content, but not on the emotional experience generated by live human interaction.

We then construct a digital twin for each participant in the AI-moderated and static conditions and evaluate its predictions against that participant's held-out responses to real-world marketing stimuli: direct mailers (images), commercials (videos), and claim-ranking (text) tasks provided by the building-products company.\footnote{As pre-registered, we did not construct digital twins in the human-moderated condition due to its much smaller sample size.} Twins built from interviews outperform demographics-only twins, but the additional richness of AI-moderated interviews does not yield more accurate predictions than static interviews. Because we collect open-ended explanations, we are able to trace prediction errors to two sources, which are neither exhaustive nor mutually exclusive: limited \emph{training data relevance}, the extent to which the information needed for the prediction task appears in the twin's training data, and limited \emph{reasoning fidelity}, the extent to which the twin's reasoning replicates the (self-reported) reasoning of the person it simulates. We find evidence for both. Consistent with reasoning fidelity relevance, humans rely more on fast, intuitive (System 1; \citealp{kahneman2011thinking}) reactions, whereas twins reason more deliberately (System 2), especially for multimodal stimuli such as images and videos. Consistent with training data relevance, adding to the training data the participant's response to one similar task improves rating predictions, suggesting a practical way to make twins more useful in market research.

\section{Methods}
\label{sec:methods}
\vspace{7pt}

\subsection{Design and Participants}
This paper reports a between-subjects study ($N=317$; 139 AI-moderated, 154 static, 24 human-moderated). All participants were recruited from Prolific and screened to be home-improvement decision-makers (homeowners, household income $\geq\$75$k, home value $\geq\$250$k, English-speaking). These screening criteria were set in collaboration with the partner brand to ensure the sample's relevance for an upscale North American company in the home-improvement category. All interviews covered the same topics and themes (See Table~\ref{tab:survey-sections} for list of different sections).

We first collected the \emph{AI-moderated} and \emph{static} conditions in the same recruitment batch. For these two conditions, we used the commercial platform developed by our AI startup partner to run the interviews. Participants who consented to participate in the study were randomly assigned to one of the conditions and directed to the platform. They first completed screening questions to verify eligibility (see ``Design and participants''). Eligible participants then completed the interview based on their assigned condition in one sitting.

In the \emph{AI-moderated} condition,%
the interviews were conducted using the platform's virtual interviewer, a turn-based, goal-conditioned conversational system.%
After each participant turn, browser-captured speech was sent to an AWS Lambda backend and transcribed using OpenAI’s gpt-4o-transcribe; the transcription request was primed with the preceding interviewer utterance and relevant study terminology. The transcript was appended to the conversation history and supplied to GPT‑5.1 through the OpenAI Chat Completions API, configured with temperature 0 and a 300-token response limit. Conditioned on the interview guide and prior dialogue, the model generated either a clarification or follow-up within the current section, or a transition to the next section. 
The virtual interviewer's response was converted to speech using the study-configured voice, OpenAI tts-1 with the alloy voice.%
Although video could be recorded, the interviewer adapted solely to the transcribed verbal content and conversation history; video, facial expressions, prosody, and other nonverbal signals were not analyzed by the language model during the interview.

In the \emph{static} condition, participants used the same platform, but were shown fixed, pre-defined questions with no adaptive follow-up or probing. In both conditions, participants recorded spoken answers to each open-ended question. 

We then recruited a separate sample for the \emph{human-moderated} condition. These interviews were conducted by GBK Collective (our market research partner), which provided trained \emph{expert} human moderators to administer the same protocol. Participants were recruited from the same online pool, but the study procedure differed slightly, due to the need to schedule live interviews.  Participants first completed a separate intake survey on Qualtrics with the same screening, demographic, and individual-difference questions as the other conditions (Sections~0, 1, and 5 --- see Table~\ref{tab:survey-sections}). Eligible participants were then messaged through Prolific to sign up for a live interview time slot, during which they completed the interview (Sections~2, 3, and 4 --- see Table~\ref{tab:survey-sections}) with the human moderator.

We made an effort to keep survey time and compensation as similar as possible across conditions: interviews in all three conditions lasted around 30 minutes. Compensation was \$10 in the AI-moderated and static conditions and \$12 in the human-moderated condition, with the additional \$2 compensating participants for completing the intake survey. The three conditions administered the \emph{same} interview and topic guide and differed only in how the interviews were moderated (please refer to the ``Interview Structure'' subsection for details and examples).

Due to the high cost of human-moderated interviews, we were unable to match its sample size to that of the other conditions. Instead, we matched conditions based on their \emph{market cost}. The market cost of running 150 AI interviews is approximately \$5k (for each participant: \$10 participant compensation for a 30-min interview + \$3.33 Prolific platform fee + \$20 AI-interview platform fee). Given the market price charged by professional human interviewers (approximately \$300 per hour, or \$150 for a 30-minute interview, plus participant compensation), \$5k translates to approximately 30 human-moderated interviews of the same length (30 min), which we preregistered. However, unexpectedly high no-show rates raised the marginal cost of that condition,\footnote{Participants were still paid an initial fee for completing the intake survey even if they later did not show up to their scheduled interviews; the interviewers still spent several minutes per no-show. Significant time was spent scheduling human-moderated interviews and following up with participants.} yielding a final sample of 24.

Following our pre-registered exclusion criteria, we excluded five AI-moderated and three static participants  who did not provide complete interview responses, either because they did not finish the interview or because technical issues prevented the platform from properly recording their responses. We excluded one additional participant from the static condition who completed the interview but did not provide meaningful answers to the open-ended blocks ($58$ words in $10$ open-ended questions). Our final sample consists of $N=139$ in the AI-moderated condition, $N=154$ in the static condition, and $N=24$ in the human-moderated condition. 

\subsection{Study Materials}\label{sec:interview-content}

Table \ref{tab:survey-sections} gives an overview of the various sections in our study. Sections 0 (screeners), 1 (Demographics), 4 (Marketing Stimuli), 5 (Individual differences) contained the same questions across the three conditions. Sections 2 (Category-specific) and 3 (Brand-specific) contained interview-style questions that were implemented differently across conditions. Web Appendix~\ref{app:interview-protocol} reports the full study materials.
\begin{table*}[h]
\TABLE
{Survey sections: content, response format, and use in twin calibration.\label{tab:survey-sections}}
{\small\hyphenpenalty=10000
\begin{tabular}{p{0.18\linewidth}p{0.37\linewidth}p{0.24\linewidth}p{0.11\linewidth}}
\toprule
Section & Content & Response format & Use in twin calibration \\
\midrule
0. Screeners & Age, homeownership, home value, household income, decision role, audio comfort, project types and description & Multiple choice & Training \\
\addlinespace
1. Demographics & Gender, ZIP code, education, employment, household composition & Multiple choice & Training \\
\addlinespace
2. Category-specific & Two open-ended blocks: Category usage, context and feelings; Purchase journey. \newline \emph{Branches:} (i)~recently replaced, (ii)~considering. & {\setlength{\fboxsep}{1.5pt}Open-ended} (differs by condition) & Training \\
\addlinespace
3. Brand-specific & Two open-ended blocks: Brand awareness; Brand perception and satisfaction. \newline \emph{Branches:} (i)~aware \& past purchaser, (ii)~aware but not a past purchaser, (iii)~not aware. & {\setlength{\fboxsep}{1.5pt}Open-ended} (differs by condition) & Training \\
\addlinespace
4. Marketing stimuli & Two mailers and two commercials (five 1--5 rating items plus an open-ended response each); two marketing-claim ranking sets (each requires ranking three claims, then an open-ended response) & Rating scales + free response & Validation (held out) \\
\addlinespace
5. Individual differences & Maximizing tendency, price consciousness, minimalism, consumer need for uniqueness & Likert scales & Training \\
\addlinespace
6. Survey experience & Enjoyment; comfort speaking and recording (AI-moderated and static conditions only) & 1--7 ratings & --- \\
\bottomrule
\end{tabular}}
{\emph{Notes:} Only Blocks~2 and 3 were open-ended and differed across interview conditions; all other blocks were administered identically. Block~4 (validation stimuli) was held out from twin construction and used only for evaluation. Block~6 was not collected in the human-moderated condition.}
\end{table*}

Section 2 and 3 each contained two blocks of open-ended questions. Section 2 focused on category-specific experiences, and Section 3 focused on brand-specific perceptions. Together, these blocks covered four topics: category usage, context and feelings; purchase journey; brand awareness; brand perception and satisfaction. For each of the four open-ended blocks, the static condition showed participants a fixed question and asked them to record a single spoken answer with no additional probing. The AI moderator was instead given an ``objective'' specifying the themes each block should cover and instructed to use their judgment to probe adaptively. The human moderators were provided with an interview guide that contained both the fixed question and moderator objective, and were given discretion to ask questions and probe naturally while covering the specified content. As Table~\ref{tab:c-example} illustrates, we designed the moderator objectives to stay as close as possible to the fixed questions, so that the conditions differed in moderation style rather than in the underlying topics. Table~\ref{tab:example-exchanges} in Web Appendix~\ref{app:example-exchanges} reproduces verbatim excerpts of an AI-moderated and a human-moderated execution of this block.

\definecolor{djgreen}{RGB}{0,120,60}
\begin{table}[h]
\TABLE
{An example of an open-ended block (purchase journey for past purchasers): fixed question versus moderator objective.\label{tab:c-example}}
{\small
\begin{tabular}{p{0.47\linewidth}p{0.47\linewidth}}
\toprule
Fixed question in Static Condition & Objective shared with Human/AI moderators \\
\midrule
Please walk us through your replacement journey from the moment you realized you needed new windows or doors to the moment you made your final choice. In your answer, please cover: \newline
\textbullet~\textcolor{teal}{What triggered the need} \newline
\textbullet~\textcolor{purple}{What research you did} \newline
\textbullet~\textcolor{orange}{Who was involved} \newline
\textbullet~\textcolor{magenta}{What options you considered} \newline
\textbullet~\textcolor{brown}{How you narrowed things down} \newline
Looking back, \textcolor{red}{how do you feel about the decision}? Share \textcolor{olive}{what mattered most} and \textcolor{djgreen}{whether you would do anything differently}. 
&
Understand the respondent's completed replacement journey from initial need to final choice: what \textcolor{teal}{triggered the need}; what \textcolor{purple}{research they did}, \textcolor{orange}{who was involved}, what \textcolor{magenta}{options they considered}, and how they \textcolor{brown}{narrowed them down}; \textcolor{olive}{what mattered most in the decision}; and how they \textcolor{red}{feel about the outcome} and whether they would \textcolor{djgreen}{do anything differently}. \\
\bottomrule
\end{tabular}}
{\emph{Notes:} The three conditions shared the same question objectives but differed in how respondents were interviewed: In the static condition, respondents were shown the fixed question and recorded a single spoken answer with no probing. In the AI condition, the moderator was given that block's objective and probed adaptively to address that objective. In the human condition, following industry best practices, the interviewer worked from both the fixed question and the objective and used their discretion to ask questions and probe naturally while covering the specified content. Each colored phrase marks a theme; the purchase journey block for past purchasers comprised eight themes (colors were not shown to participants or moderators).}
\end{table}

Section 4 presented six real-world marketing-stimulus tasks supplied by the partner brand: two direct-mail images, two commercial videos, and two sets of three marketing claims. Due to confidentiality concerns, we are not able to share the exact marketing stimuli tested in the survey. Participants evaluated the same stimuli using fixed questions across conditions, without adaptive probing.

For each mailer and commercial, participants rated five statements: “This [mailer/commercial] grabs my attention,” “The main message of this [mailer/commercial] is clear,” “I like this [mailer/commercial] overall,” “This [mailer/commercial] makes me want to learn more about [the partnering brand],” and “This [mailer/commercial] feels relevant to me and my needs” ($1$ = strongly disagree, $5$ = strongly agree). Each stimulus was followed by an open-ended question: “What stood out to you most about this [mail piece/commercial], and what (if anything) could be improved?”

The two mailers differed in visual style and content. The first featured a handwritten-style design with highlighted discounts, while the second used a brochure-style layout with product photographs and promotional offers. The two commercials each lasted approximately 30 seconds. The first emphasized customer service through testimonials about the replacement experience, while the second highlighted a limited-time financing offer alongside images of windows, doors, and home interiors.

The two claim tasks asked participants to rank three benefit claims specific to the company's products by attractiveness (\#$1$ = resonates the most, \#$3$ = resonates the least) and three claims about a particular window product by concern (\#$1$ = most concerning, \#$3$ = least concerning), respectively. For each set, participants were instructed to explain why their top-ranked selection was more attractive/concerning than the other claims.%
 Web Appendix~\ref{app:interview-protocol} (Section~4) reports the complete wording.

Participants also completed individual-difference items (Section 5) adapted from established measures: maximizing tendency (five items, adapted from \citealp{nenkov2008maximization}; e.g., “I never settle for second best”), price consciousness (five items, adapted from \citealp{lichtenstein1993price}; e.g., “I will grocery shop at more than one store to take advantage of low prices”), consumer minimalism (12 items, adapted from \citealp{wilson2021minimalism}, e.g., “I avoid accumulating lots of stuff”), and consumer need for uniqueness (12 items, adapted from \citealp{ruvio2008shortform}, e.g., “I actively seek to develop my personal uniqueness by buying special products or brands”). Maximizing tendency, price consciousness, and need for uniqueness used five-point agreement scales ($1$ = strongly disagree, $5$ = strongly agree); minimalism used a seven-point agreement scale ($1$ = strongly disagree, $7$ = strongly agree). Web Appendix~\ref{app:interview-protocol} reports the complete lists of items.

At the end of the survey, participants in the AI-moderated and static conditions answered two single-item experience questions (Section 6): “How enjoyable was it to complete this study?” ($1$ = not enjoyable at all, $7$ = very enjoyable) and “How comfortable was it to speak and record your answers?” ($1$ = not comfortable at all, $7$ = very comfortable). We analyzed the two ratings separately. These questions were not administered in the human-moderated condition.

\subsection{Interview Process and Data-Quality Measures}\label{sec:process-measures}

\noindent We characterized each interview along four families of measures, summarized in Table~\ref{tab:measures}. The volume, richness, affect, and prosody measures were computed on \emph{interviewee-only} speech from the four open-ended blocks in the category and brand-specific sections (Sections 2 and 3). We transcribed each interview with Whisper timestamps \citep{radford2023whisper} and matched each Automatic Speech Recognition (ASR) segment to the known-speaker transcript by token overlap. For participant-level interview outcomes, we compare independent condition means using two-sided Welch's $t$-tests, which allow unequal variances and sample sizes. 

\begin{table}[h]
\TABLE
{Interview-process and data-quality measures.\label{tab:measures}}
{\small
\begin{tabular}{p{0.26\linewidth}p{0.64\linewidth}}
\toprule
Measure & Definition \\
\midrule
\multicolumn{2}{@{}l}{\emph{Interview volume}} \\
Participant speaking time & Total minutes of participant speech across the four open-ended blocks \\
Overall word count & Total participant words across the four blocks \\
Participant turns & Total number of participant-speaking turns \\
\addlinespace
\multicolumn{2}{@{}l}{\emph{Richness and depth per open-ended block}} \\
Breadth & Number of the block's predefined themes the participant covers, coded by an LLM against the per-block codebook (see full list of themes in Table~\ref{tab:block-codebooks} in Web Appendix~\ref{app:llm-coding}) \\
Mean depth & Average elaboration across present themes only ($1$ = mention only, $4$ = rich, detailed elaboration). Absent themes receive $0$ at the coding stage and are excluded from this average; a block with no present themes receives $0$ (examples in Web Appendix~\ref{app:llm-coding}) \\
\addlinespace
\multicolumn{2}{@{}l}{\emph{Speech-based affect}} \\
Valence & How positive versus negative the participant sounds \\
Arousal & How activated or energized the participant sounds \\
Dominance & How assertive the participant sounds \\
\addlinespace
\multicolumn{2}{@{}l}{\emph{Prosody}} \\
Voiced fraction & Share of frames with detected pitch \citep{mauch2014pyin} \\
Speaking rate & Participant words divided by voiced seconds, multiplied by 60 (words per voiced minute) \\
Pause fraction & Share of frames below 10\% of the recording's mean RMS energy \\
\bottomrule
\end{tabular}}
{}
\end{table}

\emph{Interview volume.} We first assessed how much speech each condition elicited and how interactive the exchange was. For each participant, speaking time equals the total durations of participant-attributed speech segments across the four open-ended blocks from Sections 2 and 3, reported in minutes. Word count is the number of whitespace-separated words in the corresponding participant text, summed across the blocks. A participant turn is a distinct response to a fixed question or moderator utterance; multiple transcription segments within that response count as one turn. We excluded moderator speech, closed-ended items, and validation responses from these measures.

\emph{Richness and depth per open-ended block.} Volume alone does not show whether an interview is substantively rich. A strong qualitative interview should cover relevant topics and explore them in depth. We therefore coded each open-ended block for breadth, the number of distinct themes a participant mentions regardless of richness, and depth, how extensively each theme is discussed. To this end, we decomposed each block's moderator objective into a fixed set of codable themes. For example, the purchase journey example in Table~\ref{tab:c-example} contains eight themes (e.g., what triggered the need to purchase). 

We coded each participant's responses to each block against this theme codebook using LLM coding (\texttt{gpt-4o}) \citep{blanchard2025newtools}.  For each participant, LLM coders rated each theme on a 0--4 scale: 0 = the theme is not meaningfully present, 1 = mention only -- present but only as a brief mention or label, 2 = minimal elaboration -- present with one layer of development, 3 = clear elaboration -- clearly developed with at least two kinds of substantive support, 4 = rich elaboration -- developed in a rich, multi-part way with several specifics and strong sense of reasoning, process, or consequences. For example, within the Decision criteria theme, “Cost and quality would be the two main things” received a rating of 1; “The price would be the most important, and quality, making sure they're going to last a long time, maybe energy savings, but that's about it” received a rating of 3; and “the main thing that matters to us is the reputation of the company […] Other top things I look for in the actual doors and windows is the warranty […] I want to make sure that I'm getting a good quality product, even if I have to pay more for it. I want to make sure that it will withstand the hurricane force winds and the rain and everything” received a rating of 4.

To improve coding reliability, we used five independent LLM coders. A theme was counted as present if at least three coders rated it above 0. For each theme counted as present, depth was defined as the median score among coders who rated that theme above 0. For each block, we calculated breadth as the number of themes present and mean depth as the average depth across themes that were present (0 if none were present). We report results separately by interview block because blocks differ in their number of themes (from 3 to 9). Web Appendix~\ref{app:llm-coding} provides the full list of themes, coding procedures, and worked examples.

\emph{Speech-based affect.} How a participant sounds carries information about emotional engagement that the transcript alone cannot capture. We scored each participant's audio recording on valence, arousal, and dominance, the standard dimensional representation of emotion \citep{mehrabian1974approach, mehrabian1996pad, russell1980circumplex}. We used a wav2vec2 speech-emotion model fine-tuned on the MSP-Podcast corpus \citep{lotfian2019msppodcast}, applied it to 8-second audio windows, and averaged the window-level scores into one interview-level score weighted by window duration \citep{wagner2023dawn}.

\emph{Prosody.} We also examined acoustic delivery, which captures how fluently and continuously participants speak. We extracted three interpretable prosodic features from the audio recordings: voiced fraction, speaking rate, and pause fraction. For each participant, voiced fraction is the number of frames with detected pitch divided by the total number of analyzed frames. Speaking rate is $60\times W/V$, where $W$ is the participant word count for the analyzed audio and $V$ is the duration of voiced frames in seconds. Pause fraction is the number of frames whose root-mean-square (RMS) energy falls below $10\%$ of that recording's mean RMS energy, divided by the total number of frames. These measures were computed with the standard \texttt{librosa} toolkit \citep{mcfee2015librosa}, using the pYIN estimator to detect voiced frames \citep{mauch2014pyin}.

\subsection{Customer Needs as Managerial Insight}
\label{sec:insight-methods}
Beyond how much participants say and how they sound, what practitioners ultimately value from qualitative interviews is the set of \emph{customer needs} they reveal \citep{berger2020uniting}. In scoping this study with our industry collaborators, the central question was whether an AI moderator surfaces the same needs as a skilled human interviewer, since these needs inform downstream product, positioning, and messaging decisions. We therefore treat elicited customer needs as the unit of managerial insight.

We define a customer need following \citet{timoshenko2026} and building on the Voice of the Customer tradition of \citet{griffin1993}: an underlying benefit the customer wants, articulated at a decision-relevant level and grounded in the respondent's own words, rather than a product feature, solution, or opinion. This corresponds to the most granular \emph{tertiary} need in \citet{griffin1993}. We adopted the LLM-based VOC pipeline of \citet{timoshenko2026}, which builds on machine-assisted need identification from consumer text \citep{timoshenko2019identifying}. For each interview, we used the respondent's answers to the four open-ended blocks and extracted needs with one \texttt{gpt-4o} pass per interview (\texttt{temperature}$=0$, prompt in Web Appendix~\ref{app:cn-prompt}). We then winnowed overlapping mentions into a master codebook by embedding each mention with \texttt{all-MiniLM-L6-v2} \citep{reimers2019sentencebert, wang2020minilm} and clustering the embeddings using cosine distance $<.60$. We treated each cluster as one unique need, with its modal phrasing as the exemplar \citep{depaoli2025codebook,tamkin2024clio}. This procedure yielded $252$ unique needs across all interviews. We then used the codebook to calculate the number of needs that emerged in each condition and their corresponding frequency. A condition's unique-need count is the number of distinct master-codebook needs recovered at least once in that condition; repeated mentions do not increase coverage. For a set of $n$ interviews, coverage is the size of the union of their recovered needs.%

\subsection{Twin Construction}\label{sec:twin-construction}

For each participant, we constructed a digital twin by conditioning an LLM (\texttt{gemini-3.1-flash-lite-preview}) on the participant's persona through in-context learning: we provided the participant's training data as context and queried the model on the six marketing-stimulus tasks (Table~\ref{tab:survey-sections}, Section~4). The stimuli and response options were supplied to the twin. We used the Gemini family because the validation stimuli are multimodal: the mailers are images and the commercials are videos, which Gemini models can process natively. We compared two twin specifications: \emph{Demographics-only} (Demo), which conditioned on demographics alone (Table~\ref{tab:survey-sections}, Sections 0 and 1), and \emph{Full}, which also included the interview transcript (Table~\ref{tab:survey-sections}, Sections 2, 3) and individual-differences scales (Table~\ref{tab:survey-sections}, Section 5). The persona system prompt and the task prompts are reported in Web Appendix~\ref{app:twin-prompts}. To test robustness to model choice, we replicated twin construction using identical prompts with \texttt{gpt-5-mini}, a model from a different family. Because \texttt{GPT} does not support video inputs, the replication covered only the claims and mailer tasks. Results are largely consistent (Web Appendix~\ref{app:model-robust}).

\subsection{Twin Evaluation and Metrics}\label{sec:validation}
We evaluated twin predictive validity on the marketing stimuli presented in Section~4 of the survey (see Table~\ref{tab:survey-sections}). Each participant's own ratings, claim rankings, and open-ended explanations served as ground truth. Table~\ref{tab:twin-metrics} summarizes the response formats and corresponding performance metrics. We describe the four types of dependent variables specified in the pre-registration: ratings accuracy (for mailers and commercials), ratings Kendall's $\tau_b$ (for mailers and commercials), rankings Kendall's $\tau$ (for claim rankings), and open-ended responses cosine similarity (for all stimuli). All measures are calculated separately for each interview condition and persona specification.

\textit{Mailers and commercials.} Participants evaluated two direct mailers and two commercials, each using five questions using 1--5 scales. We evaluated twin predictions using two measures. First, for each rating question, we calculated the mean absolute error (MAE) between the twin-predicted and human ratings across participants and transformed it into accuracy as $1-\mathrm{MAE}/4$ (4 is the maximum possible difference on a 1--5 scale). We then averaged accuracy across the five questions for each stimulus. Accuracy ranges from 0, when human and twin ratings are at opposite ends of the scale, to 1, when they are identical.

Second, for each rating question, we calculated Kendall's $\tau_b$ between human and twin ratings across participants. This measures whether twins preserve the ordering of participants' responses to that question independent of their absolute level. We calculated $\tau_b=(C-D)/\sqrt{(C+D+U_H)(C+D+U_T)}$, where $C$ and $D$ count concordant and discordant pairs of participants, and $U_H$ and $U_T$ count pairs tied only in the human or twin ratings, respectively. Values range from -1 (completely reversed ordering of participants) to 1 (identical ordering of participants); 0 indicates no relationship between the two orderings. Kendall's $\tau_b$ is undefined when either human or twin ratings are constant across participants, resulting in a zero denominator. To ensure consistent comparisons across conditions and bootstrap iterations (we drew 1,000 bootstrapped samples of the same size by resampling with replacement in order to obtain confidence intervals), we excluded any rating item for which $\tau_b$ was undefined in the observed sample or in any of the bootstrap draws in any condition. This excluded 8 of the 20 rating items. This exclusion only applies to the $\tau_b$ measure. For this reason, and for parsimony, we report Kendall's $\tau$-b in the Web Appendix~\ref{app:bootstrap-inference}.

\textit{Marketing claims.} Participants completed two claim-ranking tasks, each requiring them to rank three marketing claims. For each task, we compared each participant's ranking with their twin's predicted ranking using Kendall's $\tau=(C-D)/3$, where $C$ and $D$ count concordant and discordant claim pairs. Identical rankings yield 1 and reversed rankings yield $-1$. For each participant, we averaged the performance across the two ranking tasks. 

\textit{Open-ended explanations.} Participants provided an open-ended explanation for each of the six stimuli. We embedded each human response and its corresponding twin response using \texttt{gemini-embedding-2} and calculated their cosine similarity, which ranges from $-1$ to 1. Higher cosine similarity indicates greater semantic similarity between the responses.

\textit{Pooled measures.} Our primary dependent variables are three pooled measures derived from the metrics above. For each participant, we averaged their stimulus-level accuracy across the four rated stimuli (two mailers and two commercials) to obtain Accuracy${_{all}}$ (ratings accuracy), the $\tau$ scores across the two ranking tasks to obtain $\tau{_{all}}$ (ranking accuracy), and the six open-ended explanation similarity scores to obtain Cosine$_{all}$ (open-ended similarity). Each stimulus receives equal weight within its pooled measure.

\begin{table}[h]
\TABLE
{Validation stimuli, response formats, and performance metrics.\label{tab:twin-metrics}}
{\small
\begin{tabular}{p{0.22\linewidth}p{0.30\linewidth}p{0.38\linewidth}}
\toprule
Stimulus (count) & Response format & Performance metric(s) \\
\midrule
Images: direct mailers ($\times 2$) & Five 1--5 scale items, plus one open-ended explanation & Accuracy ($1-|H-T|/4$) and Kendall's $\tau$-b on the scale items; cosine similarity on the open-ended response \\
\addlinespace
Videos: commercials ($\times 2$) & Five 1--5 scale items, plus one open-ended explanation & Accuracy ($1-|H-T|/4$) and Kendall's $\tau$-b on the scale items; cosine similarity on the open-ended response \\
\addlinespace
Text: social media claim ($\times 2$ sets) & Ranking of three claims, plus one open-ended explanation & Kendall's $\tau$ on the ranking; cosine similarity on the open-ended response \\
\bottomrule
\end{tabular}}
{\emph{Notes:} $H$ is the human response and $T$ the twin prediction. 
}
\end{table}

\subsection{Understanding Prediction Error: Training Data Relevance vs. Reasoning Fidelity}\label{sec:gaps-method}
Twin prediction errors can arise from at least two conceptually distinct sources. First, the training data may be too far removed from the prediction task, i.e., the twin was not given enough relevant data to predict a particular response. Second, the (self-reported) reasoning from the digital twin may be different from that of the human they are supposed to mimic, i.e., the human and the twin ``think'' differently about the prediction task. 

To measure training data relevance, we used an LLM judge (\texttt{gpt-4o-mini}, temperature~$0$) to code the extent to which the human's explanation of their heldout choice can be traced to themes in the interview. Specifically, the LLM judge received the participant's interview transcript together with that participant's own open-ended explanation of their heldout response, and rated on a 0--4 scale how much the participant's reasoning in their explanation of response to the validation question can be traced back to specific statements made in the interview ( 0 = no reasoning point traces to an interview statement; 4 = every reasoning point traces to an interview statement). Note that the twin's
output played no role in this measure: the measure captured only whether the information the human relied upon when evaluating the holdout stimulus was present in the training data. 

To measure reasoning fidelity, we also instructed an LLM to assess the extent to which the twin reasons similarly to the
human, based on their explanations of their response to each heldout question: the same LLM judge (\texttt{gpt-4o-mini}, temperature~$0$) compared the twin's predicted explanation with the human's actual explanation for the same stimulus, rating on a 0--4 scale ( 0 = the twin reasons entirely differently from the human; 4 = the twin reasons with the same logic/priorities as the human). Web Appendix~\ref{app:gap-prompts} reports the coding prompts and examples.

We note that training data relevance and reasoning fidelity are not orthogonal concepts. In particular, gaps in self-reported reasoning may be in part the result of the training data not being completely relevant to the task. Nevertheless, regressing the predictive validity of digital twins on training data relevance and reasoning fidelity gives insight into promising areas of improvement for digital twins.

\subsection{Similar-Task Calibration Ladder}
We also ask whether observing a person's response to a similar task can reduce the data gap and thus improve the twin's prediction. Specifically, we tested whether adding the person's \emph{choice} (rating/ranking) and open-ended \emph{response} for the \emph{other} marketing stimulus of the same type improves prediction on the target stimulus (e.g., adding the responses to one commercial as training data to predict the responses to the other commercial). Web Appendix~\ref{app:ladder-prompts} shows the prompt inserts.

\section{Results: AI-Moderated Interviews as a Source of Consumer Insight}
We first present results related to our first research question: how AI-moderated interviews compare to static interviews and human-moderated interviews in collecting qualitative data and insights.  
\subsection{Interview Process and Data Quality}
\label{sec:interview-process-results}

\label{sec:volume}
We first asked whether AI moderation produces interview data comparable in volume to a human moderator, and how it compares with a static interview. Using the interview-process and data-quality measures described in the ``Interview Process and Data-Quality Measures'' subsection, Table~\ref{tab:quality-desc} reports interview volume and participant experience across conditions. Relative to static, AI moderation elicited substantially more participant speech: participants spoke nearly twice as long ($8.1$ vs.\ $4.3$ minutes), produced more words ($952$ vs.\ $609$), and took more turns ($30.5$ vs.\ $8.3$;\footnote{As Table~\ref{tab:survey-sections} shows, the static condition branches based on participants' prior purchase experience and brand awareness, so the average number of turns is not a whole number.} all $p<.001$). Human moderation produced the highest speech volume, with significantly longer speaking time and higher word count than AI moderation ($11.3$ vs.\ $8.1$ minutes, $p=.003$; $1{,}404$ vs.\ $952$ words, $p=.006$). However, AI and human moderation generated a similar number of participant turns ($30.5$ vs.\ $32.4$, $p=.238$), suggesting a comparable level of interaction. Finally, compared to static, the additional interactions resulting from adaptive questioning in the AI condition did not diminish participant experience: enjoyment and comfort were statistically similar between AI moderation and static. Web Appendix~\ref{app:word-count-breakdown} breaks down the word counts by open-ended block.

\begin{table}[h]
\TABLE
{Interview process and participant experience by condition.\label{tab:quality-desc}}
{\footnotesize
\resizebox{\linewidth}{!}{%
\begin{tabular}{lrrrrr}
\toprule
Condition & Speaking time (min.) & Word count & Speaking turns & Enjoyment (1--7) & Comfort (1--7) \\
\midrule
AI-moderated ($N{=}139$)    & $8.1$ $(5.2)$  & $952.3$ $(632.8)$  & $30.5$ $(4.4)$ & $5.65$ $(1.29)$ & $6.37$ $(1.04)$ \\
Static ($N{=}154$)    & $4.3$ $(2.5)$  & $609.2$ $(336.9)$  & $8.3$ $(1.0)$  & $5.59$ $(1.20)$ & $6.27$ $(1.12)$ \\
Human-moderated ($N{=}24$)  & $11.3$ $(4.4)$ & $1403.6$ $(702.7)$ & $32.4$ $(7.4)$ & --             & --             \\
\addlinespace
$p$ [$d$] (AI-mod. vs.\ Static) & $<.001$ [$0.95$] & $<.001$ [$0.69$] & $<.001$ [$7.22$] & $.699$ [$0.05$] & $.393$ [$0.10$] \\
$p$ [$d$] (Human-mod. vs.\ AI-mod.)  & $.003$ [$0.62$]   & $.006$ [$0.70$]   & $.238$ [$0.38$]        & --     & --     \\
\bottomrule
\end{tabular}}}
{\emph{Notes:} Speaking time, word count, and speaking turns were calculated based on the open-ended questions from Sections 2-3. Enjoyment and comfort are self-reported scores (1 = not enjoyable/comfortable at all; 7 = very enjoyable/comfortable) and were not collected in the human-moderated condition. Entries are participant-level means with sample SDs in parentheses. $p$ values are from two-sided Welch independent-samples $t$-tests of the indicated condition means. Cohen's $d$ is the first named condition mean minus the second, divided by the sample-size-weighted pooled within-condition SD. Comparison cells report $p$ followed by Cohen's $d$ in brackets.}
\end{table}

\subsubsection{Interview breadth and depth}
\label{sec:richness}
Table~\ref{tab:quality-block} reports interview richness, reflected by breadth of theme coverage (number of themes covered out of a set maximum, see the ``Interview Process and Data Quality'' subsection) and mean depth of responses. Relative to static interviews, AI-moderated interviews covered more themes in three of the four blocks: purchase journey ($7.29$ vs.\ $5.90$, $p<.001$), brand awareness ($3.32$ vs.\ $1.73$, $p<.001$), and brand perception and satisfaction ($5.88$ vs.\ $5.52$, $p<.001$). AI moderation also produced greater mean depth in purchase journey ($2.56$ vs.\ $2.31$, $p<.001$) and brand perception and satisfaction ($2.71$ vs.\ $2.55$, $p<.001$). The short category usage, context and feelings block, which was limited to three turns (other blocks did not have such limit), was the exception, showing no difference in either breadth or depth.

\begin{table}[h]
\TABLE
{Interview data quality by open-ended block.\label{tab:quality-block}}
{\footnotesize
\resizebox{\linewidth}{!}{%
\begin{tabular}{llrrr}
\toprule
Block & Condition & Word count & Breadth & Mean depth \\
\midrule
\multirow{5}{*}{\shortstack[l]{Category usage, context and feelings\\ \footnotesize(max. breadth = 3)}}
  & AI-moderated    & $99.7$ $(69.5)$   & $2.94$ $(.25)$ & $2.51$ $(.43)$ \\
  & Static    & $111.1$ $(84.1)$  & $2.94$ $(.24)$ & $2.53$ $(.45)$ \\
  & Human-moderated & $412.0$ $(246.3)$ & $2.92$ $(.41)$ & $2.96$ $(.45)$ \\
  & \quad $p$ [$d$] (AI-mod. vs.\ Static) & $.21$ [$-0.15$] & $.82$ [$-0.03$] & $.74$ [$-0.04$] \\
  & \quad $p$ [$d$] (AI-mod. vs.\ Human-mod.)& $<.001$ [$-2.76$] & $.83$ [$0.07$] & $<.001$ [$-1.03$] \\
\addlinespace
\multirow{5}{*}{\shortstack[l]{Purchase journey\\ \footnotesize(max. breadth = 9)}}
  & AI-moderated    & $413.6$ $(305.0)$ & $7.29$ $(.91)$ & $2.56$ $(.28)$ \\
  & Static    & $197.6$ $(123.1)$ & $5.90$ $(1.44)$ & $2.31$ $(.31)$ \\
  & Human-moderated & $414.2$ $(271.4)$ & $6.08$ $(1.61)$ & $2.46$ $(.28)$ \\
  & \quad $p$ [$d$] (AI-mod. vs.\ Static) & $<.001$ [$0.95$] & $<.001$ [$1.14$] & $<.001$ [$0.85$] \\
  & \quad $p$ [$d$] (AI-mod. vs.\ Human-mod.)& $.99$ [$0.00$] & $.001$ [$1.16$] & $.116$ [$0.36$] \\
\addlinespace
\multirow{5}{*}{\shortstack[l]{Brand awareness\\ \footnotesize(max. breadth = 4)}}
  & AI-moderated    & $116.4$ $(88.9)$  & $3.32$ $(.80)$ & $2.25$ $(.48)$ \\
  & Static    & $80.9$ $(56.7)$   & $1.73$ $(.53)$ & $2.14$ $(.65)$ \\
  & Human-moderated & $273.8$ $(258.2)$ & $1.79$ $(.88)$ & $1.99$ $(.80)$ \\
  & \quad $p$ [$d$] (AI-mod. vs.\ Static) & $<.001$ [$0.48$] & $<.001$ [$2.37$] & $.089$ [$0.20$] \\
  & \quad $p$ [$d$] (AI-mod. vs.\ Human-mod.)& $.007$ [$-1.23$] & $<.001$ [$1.88$] & $.131$ [$0.49$] \\
\addlinespace
\multirow{5}{*}{\shortstack[l]{Brand perception and satisfaction\\ \footnotesize(max. breadth = 6)}}
  & AI-moderated    & $322.6$ $(225.2)$ & $5.88$ $(.35)$ & $2.71$ $(.33)$ \\
  & Static    & $219.6$ $(126.7)$ & $5.52$ $(.69)$ & $2.55$ $(.35)$ \\
  & Human-moderated & $303.6$ $(193.4)$ & $5.12$ $(1.51)$ & $2.38$ $(.65)$ \\
  & \quad $p$ [$d$] (AI-mod. vs.\ Static) & $<.001$ [$0.57$] & $<.001$ [$0.65$] & $<.001$ [$0.45$] \\
  & \quad $p$ [$d$] (AI-mod. vs.\ Human-mod.)& $.67$ [$0.09$] & $.023$ [$1.15$] & $.025$ [$0.83$] \\
\bottomrule
\end{tabular}}}
{\emph{Notes:} Entries are participant-level means with sample SDs in parentheses. $p$ rows report two-sided Welch independent-samples $t$-tests of equal condition means for the indicated comparisons. Cohen's $d$ is the first named condition mean minus the second, divided by the sample-size-weighted pooled within-condition SD. Comparison cells report $p$ followed by Cohen's $d$ in brackets.}
\end{table}

As a robustness check, we compared breadth and mean depth of AI-moderated and static interviews while controlling for length of each block. For breadth, we regressed the number of themes covered on condition (AI vs. static) and total word count in that block. For mean depth, we regressed average depth on condition and mean words per covered theme. After controlling for length, the AI-moderated advantage persisted for breadth in three of the four blocks  (purchase journey, brand awareness, and brand perception and satisfaction), and for mean depth in two of the four blocks (purchase journey and brand awareness; Table~\ref{tab:quality-reg} in Web Appendix~\ref{app:quality-reg}). Thus, AI moderation did not merely produce longer responses. Conditional on length, it elicited broader and, in some blocks, deeper content than static interviews. A pairwise LLM-as-a-judge robustness check ($1{,}000$ matched response pairs per open-ended block) showed the same pattern (Web Appendix~\ref{app:llm-judge}).

Compared with human-moderated interviews, AI-moderated interviews covered more themes in three of the four blocks: purchase journey ($7.29$ vs.\ $6.08$, $p=.001$), brand awareness ($3.32$ vs.\ $1.79$, $p<.001$), and brand perception and satisfaction ($5.88$ vs.\ $5.12$, $p=.023$). Mean depth was statistically comparable in the purchase journey and brand awareness blocks, higher for AI moderation in brand perception and satisfaction ($2.71$ vs.\ $2.38$, $p=.025$), and higher for human moderation in the short category usage, context and feelings block ($2.96$ vs.\ $2.51$, $p<.001$). Thus, longer human interviews did not appear to translate into systematically broader or deeper theme coverage.

\subsubsection{Speech-based affect: do participants sound different when speaking to a human vs.\ an AI?}
\label{sec:affect}

We next compared participant audio across the three interview conditions using the speech-based affect and prosody measures described in the Methods section. As Table~\ref{tab:affect} shows, the affect measures distinguished human moderation from the two automated conditions. Participants who spoke to a human moderator sounded more positive, more emotionally activated, and more assertive than participants who spoke to an AI moderator: valence ($.528$ vs.\ $.471$), arousal ($.517$ vs.\ $.359$), and dominance ($.562$ vs.\ $.449$) were all higher in the human condition ($ps<.001$). By contrast, AI and static interviews were statistically indistinguishable on all three affect dimensions.

Prosody showed a more nuanced pattern. Voiced fraction and speaking rate did not differ significantly across conditions, but AI moderation produced more pausing than both human moderation and static. This may reflect hesitation, additional thinking, or adjustment to the AI interaction. Overall, AI moderation improved interview volume and coded richness relative to static, but it did not reproduce the emotional engagement of speaking to a human moderator in real time.

\begin{table}[h]
\TABLE
{Participant speech-based affect and prosody by modality.\label{tab:affect}}
{\footnotesize
\resizebox{\linewidth}{!}{%
\begin{tabular}{lrrrrrr}
\toprule
Condition & Valence & Arousal & Dominance & Voiced frac. & Speaking rate & Pause frac. \\
\midrule
AI ($N{=}139$)           & $.471$ $(.055)$ & $.359$ $(.101)$ & $.449$ $(.080)$ & $.634$ $(.149)$ & $245.3$ $(99.3)$  & $.266$ $(.091)$ \\
Static ($N{=}154$) & $.471$ $(.056)$ & $.354$ $(.111)$ & $.447$ $(.089)$ & $.623$ $(.150)$ & $250.4$ $(130.1)$ & $.199$ $(.073)$ \\
Human ($N{=}24$)         & $.528$ $(.046)$ & $.517$ $(.048)$ & $.562$ $(.034)$ & $.610$ $(.126)$ & $265.6$ $(69.6)$  & $.209$ $(.073)$ \\
\addlinespace
$p$ [$d$] (AI-mod. vs.\ Static) & $.948$ [$-0.01$] & $.666$ [$0.05$] & $.813$ [$0.03$] & $.538$ [$0.07$] & $.707$ [$-0.04$] & $<.001$ [$0.82$] \\
$p$ [$d$] (Human-mod. vs.\ AI-mod.)  & $<.001$ [$1.08$] & $<.001$ [$1.65$] & $<.001$ [$1.50$] & $.404$ [$-0.17$] & $.227$ [$0.21$] & $.002$ [$-0.64$] \\
\bottomrule
\end{tabular}}}
{\emph{Notes:} Valence, arousal, and dominance are in $[0,1]$. Speaking rate is words per voiced minute; voiced fraction and pause fraction are proportions of analyzed frames. Entries are participant-level means with sample SDs in parentheses. $p$ rows report two-sided Welch independent-samples $t$-tests. Cohen's $d$ is the first named condition mean minus the second, divided by the sample-size-weighted pooled within-condition SD. Comparison cells report $p$ followed by Cohen's $d$ in brackets.}
\end{table}

\subsection{Customer Needs Coverage}
\label{sec:insight-results}\par Beyond transcript richness, we asked whether the interview conditions recovered the same customer-need space, and how efficiently they recovered those needs.\par 

We extracted and consolidated customer needs as described in the ``Customer Needs as Managerial Insight'' subsection, applying the same pipeline to the four open-ended blocks in all three conditions. The resulting master codebook contained $252$ unique needs. Table~\ref{tab:needs-examples} lists the most common shared needs and examples unique to each condition.\par  

\begin{table}[!htb] 
\TABLE {Most Common Shared and Unique Customer Needs Across Interview Conditions.\label{tab:needs-examples}} {\scriptsize   \begin{tabular}{p{0.86\linewidth}r}     \toprule     Customer need & \# mentions\\     \midrule     \textbf{Most Common Needs Shared Across All Conditions} &  \\     Access to affordable window and door replacement options & 59 \\ Desire for a cost-effective solution for window replacement & 43 \\ Confidence in the quality of professional installation & 41 \\ Desire for more energy efficient windows and doors & 41 \\ Confident in the longevity and quality of windows and doors & 40 \\ Trust in the brand's reputation and quality of products & 40 \\ Confident the new windows will lower my energy bills & 39 \\ Cost-efficient window and door replacements & 37 \\ Confidence in the quality and reputation of the brand & 35 \\ Assured of the cost-effectiveness of the window replacement & 31 \\ \addlinespace     \textbf{Most Frequently Mentioned Need Unique to AI-Moderated Interviews} &  \\     Preference for quality and price balance & 8 \\ \addlinespace     \textbf{Most Frequently Mentioned Need Unique to Static Interviews} &  \\  Availability of services in my location & 4 \\ \addlinespace     \textbf{Most Frequently Mentioned Need Unique to Human-Moderated Interviews} &  \\ Confident in the ability to find reliable instructional resources for DIY projects & 1 \\  \bottomrule  
\end{tabular}} {\emph{Notes:}%
A full list is available to qualified academic researchers upon request for research purposes.} \end{table}

\subsubsection{Coverage Efficiency}

 We first compared the coverage of needs across conditions at matched sample sizes. We used resampling comparisons at matched interview counts consistent with common qualitative saturation benchmarks \citep{guest2006howmany,guest2020saturation}. We report $95\%$ bootstrapped confidence intervals; the sampling scheme for each condition is specified in Web Appendix~\ref{app:coverage-matched24}. 
 
\begin{figure}[!htbp]
\FIGURE
{\includegraphics[width=.6\textwidth]{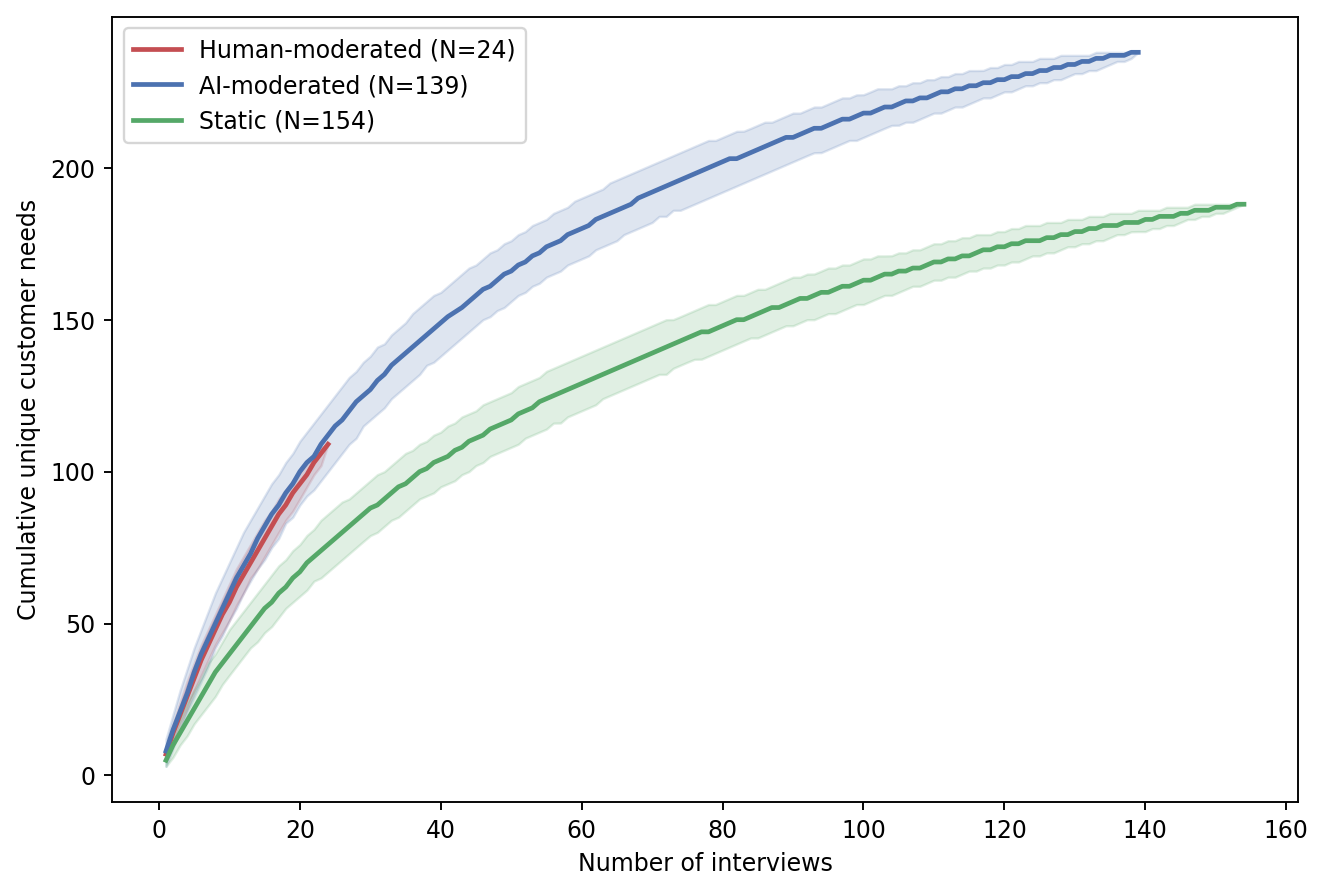}}
{Customer-need saturation curves by condition.\label{fig:coverage-curve}}
{}
\end{figure}

 Figure~\ref{fig:coverage-curve} plots customer-need saturation curves, showing how many unique needs are recovered as interviews are added. Lines show the median across random interview orderings, and shaded bands show the 5th--95th percentile range.%

Over the first 24 interviews, AI moderation and human moderation recovered a similar number of unique needs, while static accumulated needs more slowly. A resampling comparison at a matched sample size of $N=24$ showed the same descriptive pattern: AI moderation recovered $111.5$ unique needs on average (95\% resampling interval $[98.0, 125.0]$), close to human moderation ($109.0$, based on the full sample of 24 human-moderated interviews) and well above static ($75.9$, 95\% resampling interval $[65.0, 87.0]$; Web Appendix~\ref{app:coverage-matched24}, Table~\ref{tab:coverage-matched24}). 

\begin{figure}[!htbp] 
\FIGURE {\includegraphics[width=.6\linewidth]{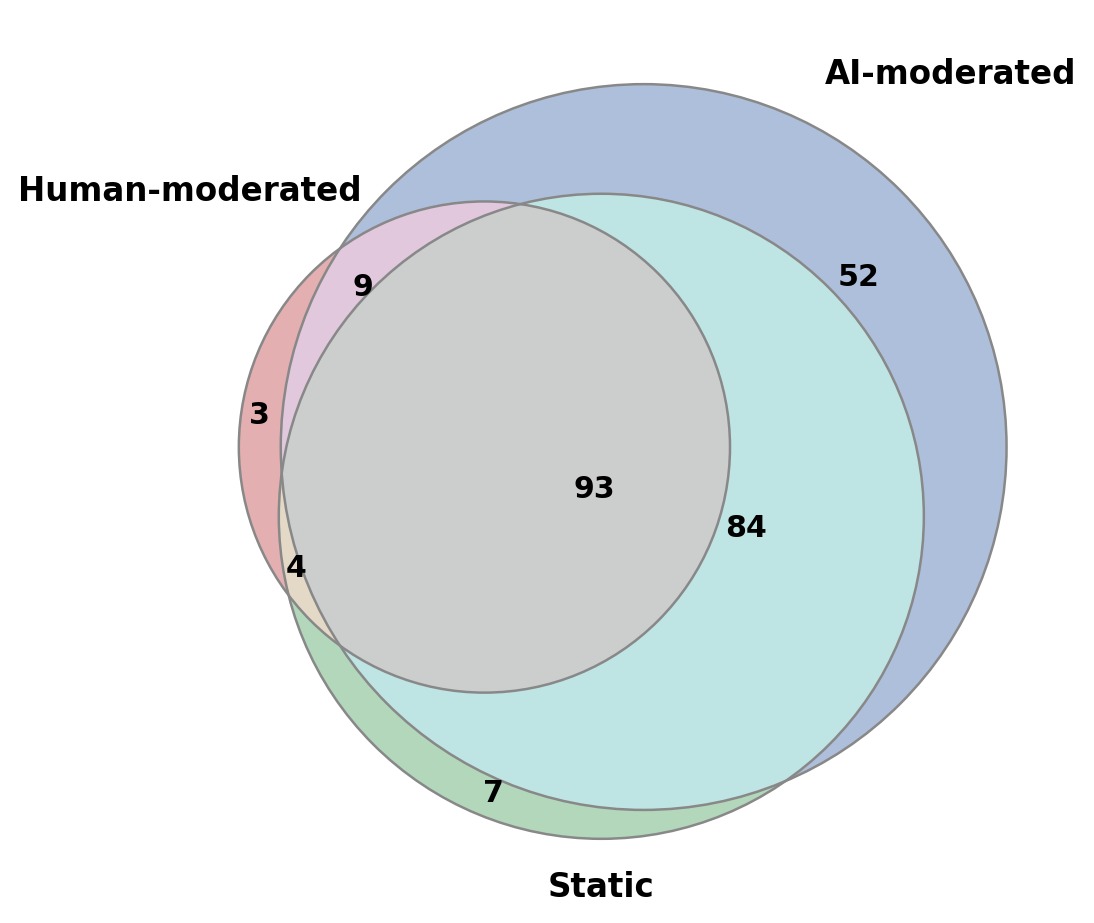}} {Customer needs shared across and unique to each interview condition.\label{fig:needs-overlap}} {} 
\end{figure}

As an additional aggregate measure, Figure~\ref{fig:needs-overlap} illustrates how the $252$ unique needs overlap across the three conditions: $93$ were recovered by all three, $52$ were unique to AI moderation, $7$ to static, and $3$ to human moderation. Table~\ref{tab:needs-examples} lists examples of the most common shared and condition-specific needs. 

\par \subsubsection{Need Elicitation Cost}\par
Because the three conditions were run under a comparable market-cost budget (\$5K), their realized full sample sizes reflect their per-interview costs. Human moderation produced many needs per interview but at roughly five times the cost of an AI-moderated interviews, so the same budget covered fewer interviews. AI moderation and static had the same per-interview cost, but AI moderation added more new needs per interview. As a result, AI moderation recovered the most unique needs under a fixed budget: $238$, compared with $188$ for static and $109$ for human moderation. Table~\ref{tab:coverage} summarizes the realized coverage and cost efficiency. The implied cost per unique need is \$36 for human moderation, \$19 for AI moderation, and \$27 for static, suggesting that AI moderation can be the most cost-efficient method for eliciting customer needs.

\begin{table}[h]
\TABLE
{Customer-need coverage and cost by condition.\label{tab:coverage}}
{\footnotesize
\begin{tabular}{lrrrr}
\toprule
Condition & Interviews & Unique needs & Cost / interview & Cost / need \\
\midrule
AI-moderated    & $139$ & $238$ & \$33.33  & \$19 \\
Static    & $154$ & $188$ & \$33.33  & \$27 \\
Human-moderated & $24$  & $109$ & \$163.33 & \$36 \\
\midrule
All combined    & $317$ & $252$ & --- & --- \\
\bottomrule
\end{tabular}}
{\emph{Notes:} Cost per need is (interviews $\times$ cost per interview) $\div$ unique needs.}
\end{table}

\section{Results: AI-Moderated Interviews as Training Data for Digital Twins}
We now address the second goal of this paper, which is to explore whether the qualitative data collected through AI-moderated interviews can be leveraged to construct digital twins capable of predicting consumer response to marketing stimuli.

\subsection{Demo vs. Full; Static vs. AI-Moderated}
\label{sec:demo-vs-full}

We compared twins built from Demographics-only personas with twins built from Full personas that included the interview transcript. Demographics-only is a strong baseline because prior work finds that synthetic personas constructed from demographic information alone can already achieve relatively high predictive accuracy \citep{peng2026funhouse}. 

We focus on three pooled measures: ratings accuracy for mailers and commercials, ranking accuracy (Kendall's $\tau$) for claims, and open-ended similarity for all stimuli. Web Appendix~\ref{app:bootstrap-inference} reports across-participant ordinal agreement on scale-rating questions using Kendall's $\tau$-b. Given the small sample size of the human-moderated condition and the high cost of collecting human-moderated interviews at scale required for digital-twin training (i.e., this is not a managerially relevant benchmark), we focus only on the AI-moderated and Static conditions in this section, as pre-registered. 

Table~\ref{tab:pooled-profile} shows that adding interview transcripts generally improved prediction quality. In the Static condition, Full personas outperformed Demo personas on all three pooled measures. In the AI-moderated condition, Full personas significantly improved open-ended similarity ($p<.001$), with a marginal gain in ratings accuracy ($p=.086$) and a positive but non-significant gain in claim-ranking accuracy ($p=.210$). Thus, interviews contain predictive signal beyond demographics alone.

\begin{table}[!htbp]
\TABLE
{Prediction performance by persona type and interview condition.\label{tab:pooled-profile}}
{\small\renewcommand{\arraystretch}{1.05}
\begin{tabular}{@{}llrrr@{}}
\toprule
    & & \shortstack{Ratings accuracy\\{\footnotesize Mailers and}\\{\footnotesize commercials}} & \shortstack{Ranking\\accuracy ($\tau$)\\{\footnotesize Claims}} & \shortstack{Open-ended\\similarity\\{\footnotesize All stimuli}} \\
\midrule
\multicolumn{5}{l}{\emph{Cell means}} \\
\multirow{2}{*}{\quad Static}
 & Demo & $.726$ & $.260$ & $.734$ \\
 & Full & $.742$ & $.344$ & $.758$ \\
\addlinespace
\multirow{2}{*}{\quad AI-moderated}
 & Demo & $.731$ & $.314$ & $.723$ \\
 & Full & $.743$ & $.345$ & $.744$ \\
\midrule
\multicolumn{5}{l}{\emph{Full vs.\ Demo, within condition ($p$ [$d_z$])}} \\
  Static &Full vs.\ Demo & $.013$ [$0.203$] & $<.001$ [$0.285$] & $<.001$ [$1.497$] \\
  AI-moderated & Full vs.\ Demo& $.086$ [$0.147$] & $.210$ [$0.107$] & $<.001$ [$1.301$] \\
\addlinespace
\multicolumn{5}{l}{\emph{AI-moderated vs.\ Static, within persona ($p$ [$d$])}} \\
 AI-moderated vs.\ Static& Demo & $.546$ [$0.071$] & $.290$ [$0.124$] & $.003$ [$-0.359$] \\
 AI-moderated vs.\ Static& Full & $.888$ [$0.016$] & $.981$ [$0.003$] & $.001$ [$-0.395$] \\
\addlinespace
\multicolumn{2}{l}{\shortstack[l]{Condition $\times$ persona\\interaction ($p$ [$d$])}} & $.687$ [$-0.047$] & $.123$ [$-0.181$] & $.120$ [$-0.184$] \\
\bottomrule
\end{tabular}}
{\emph{Notes:} Comparisons report two-sided $p$ values. Full--Demo comparisons use paired $t$-tests; AI-moderated--Static comparisons use Welch's $t$-tests. The interaction is a Welch test comparing participants' Full--Demo differences between conditions. Cohen's $d$ uses the sample-size-weighted pooled within-group SD; paired $d_z$ uses the SD of within-participant differences. Signs follow the displayed subtraction; interaction $d$ compares Full--Demo gains using their pooled within-group SD. Comparison cells report $p$ followed by the effect size in brackets: $d_z$ for Full--Demo comparisons and $d$ for between-condition comparisons and the interaction.}
\end{table}

However, we found no evidence that AI-moderated interviews produced more predictive twins than Static interviews on ratings accuracy or claim-ranking accuracy. Surprisingly, Static actually yielded slightly higher open-ended similarity under both Demo ($.734$ vs.\ $.723$, $p=.003$) and Full ($.758$ vs.\ $.744$, $p=.001$). None of the condition-by-persona interactions is statistically significant (all $p\geq .1$), suggesting no clear difference in the incremental predictive value of interview transcripts across conditions. Taken together, these results suggest that interview data can improve digital-twin predictions, but the greater richness of AI-moderated interviews does not translate into a demonstrated predictive advantage over Static interviews.

\subsection{Understanding Prediction Error: Training Data Relevance vs.\ Reasoning Fidelity}
\label{sec:gaps}

To better understand the prediction errors of digital twins, following the framework of the section ``Understanding Prediction Error: Training Data Relevance vs. Reasoning Fidelity,'' we regressed twins' ratings accuracy, ranking accuracy, and open-ended similarity on training data relevance (an LLM-coded 0--4 rating of the extent to which the human's open-ended explanation of their holdout response can be traced back to the interview) and reasoning fidelity (an LLM-coded 0--4 rating of how closely the twin's reasoning expressed in their open-ended explanation of the holdout response matches the human's). We included stimulus fixed effects and cluster standard errors by participant to account for dependence across repeated stimulus responses from the same participant. 

As Table~\ref{tab:gaps} shows, training data relevance positively predicted all three outcomes: ratings accuracy (\textit{b} = $0.029$, $p=.005$), ranking accuracy (\textit{b} = $0.117$, $p=.025$), and open-ended similarity (\textit{b} = $0.034$, $p<.001$). Reasoning fidelity served as a positive and significant predictor of ratings accuracy (\textit{b} = $0.085$, $p<.001$) and ranking accuracy (\textit{b} = $0.338$, $p<.001$), but did not significantly predict open-ended similarity (\textit{b} = $0.006$, $p=.210$). The latter finding is somewhat unexpected, as both reasoning fidelity (LLM-coded) and open-ended similarity (embedding-based) assess agreement between human and twin explanations, albeit using different theoretical constructs and measurement approaches. This pattern suggests that prediction error is not only about whether relevant information is present in the interview. Even when such information is available, twins' quantitative predictions are more accurate when their open-ended reasoning about the stimulus resembles the human's reasoning. This suggests that while some gains in predictive validity may be expected from providing training data that is more relevant to the task the twin is asked to simulate, future research may also focus on aligning the reasoning process of twins with that of humans.

\begin{table}[h]
\TABLE
{Predictive performance regressed on training data relevance and reasoning fidelity (Full personas, AI-moderated condition).\label{tab:gaps}}
{\footnotesize
\begin{tabular}{lccc}
\toprule
& \shortstack{Ratings\\accuracy} & \shortstack{Ranking\\accuracy ($\tau$)} & \shortstack{Open-ended\\similarity} \tabularnewline
\midrule
Training data relevance
& $.029^{**}$ & $.117^{*}$ & $.034^{***}$ \tabularnewline
& $(.010)$ & $(.052)$ & $(.004)$ \tabularnewline
Reasoning fidelity
& $.085^{***}$ & $.338^{***}$ & $.006$ \tabularnewline
& $(.017)$ & $(.061)$ & $(.004)$ \tabularnewline
\midrule
Stimulus FE & Yes & Yes & Yes \tabularnewline
SE clustered by participant & Yes & Yes & Yes \tabularnewline
$R^{2}$ & $.19$ & $.20$ & $.35$ \tabularnewline
Observations & $555$ & $276$ & $831$ \tabularnewline
\bottomrule
\end{tabular}}
{\emph{Notes:} Each observation is a participant--stimulus response. The dependent variables are ratings accuracy (the mean of $1-|H-T|/4$ across five questions) for mailers and commercials, ranking accuracy (Kendall's $\tau$) for claim sets, and open-ended cosine similarity for all six stimuli. Models are OLS with stimulus fixed effects; participant-clustered standard errors are in parentheses. Significance symbols use two-sided asymptotic Wald $z$-tests of coefficients against zero. $^{\dagger}p<.10$, $^{*}p<.05$, $^{**}p<.01$, $^{***}p<.001$}
\end{table}

We next explored some factors that may affect reasoning fidelity. Prior work suggests that digital twins can be overly analytical, or ``hyper-rational'' \citep{goli2024frontiers,mei2024turing,peng2026funhouse}. This matters for marketing stimuli, especially images and videos, where humans often respond quickly and affectively while twins may give more deliberate explanations. To measure this, we used an LLM judge (\texttt{gpt-4o-mini}, temperature~$0$) to code each open response, in the spirit of verbal protocol analysis \citep{ericsson1980verbal,ericsson1993protocol,nisbett1977telling}, on a scale from $0$ = intuitive/experiential (System~1) to $4$ = analytical/deliberative (System~2) \citep{kahneman2011thinking}. Each response was scored in a separate judge call, with no indication of whether the text came from the human or the twin; the coding prompt is in Web Appendix~\ref{app:gap-prompts}. The \emph{reasoning-style gap} for a response pair is the absolute difference $|$Twin $-$ Human$|$ on this scale.

Table~\ref{tab:sys12} shows that twins were substantially more analytical than humans for Mailers ($2.18$ vs.\ $1.31$, $p<.001$) and Commercials ($2.23$ vs.\ $1.60$, $p<.001$). For Claims, the difference was much smaller, though still statistically significant ($2.75$ vs.\ $2.50$, $p<.001$). A separate response-focus coding reported in Web Appendix~\ref{app:response-focus} suggests humans were more likely to react to what they directly see or hear in the stimulus, whereas twins were more likely to draw on person-level context from the interview.

Reasoning fidelity and reasoning-style gap are two measures of the difference between the twin's and the human's open-ended reasoning about the same stimulus, yet empirically they were only weakly correlated ($r = -.23$, $-.15$, and $-.25$ for Claims, Mailers, and Commercials). The tendency of twins to reason more analytically than humans therefore accounts for only part of the divergence between twin and human reasoning, suggesting that other differences between the two remain to be identified and addressed. Because both are measures of the twin--human reasoning difference, we did not enter them jointly in our regressions. Web Appendix~\ref{app:gaps-alt} reports separate task-family regressions with the reasoning-style gap in place of reasoning fidelity: the gap significantly predicted lower accuracy for Claims, with small and nonsignificant coefficients for Mailers and Commercials. This further confirms that the twins' tendency to reason more analytically only accounts for a modest portion of the divergence between twins and humans.

\begin{table}[h]
\TABLE
{Open-response style on a $0$--$4$ scale ($0=$ intuitive/System~1,
  $4=$ analytical/System~2).\label{tab:sys12}}
{\small
  \begin{tabular}{lrrrrr}
    \toprule
    Task & $n$ & Human & Twin & $\Delta$ & $p$ [$d_z$] \\
    \midrule
    Claims & $139$ & $2.50$ & $2.74$ & $+.25$ & $<.001$ [$0.387$] \\
    Mailers & $139$ & $1.31$ & $2.18$ & $+.87$ & $<.001$ [$1.575$] \\
    Commercials & $139$ & $1.60$ & $2.23$ & $+.63$ & $<.001$ [$1.108$] \\
    \bottomrule
  \end{tabular}}
{\emph{Notes:} AI-moderated participants only. Human and Full-persona twin scores were averaged across matched stimuli within each task family for each participant. Responses were independently scored by an LLM judge blinded to their source. Entries are participant-level means; $\Delta$ is Twin $-$ Human. $p$ values are from two-sided paired $t$-tests. Paired Cohen's $d_z$ is the mean participant-level Twin--Human difference divided by the SD of those differences. The $p$ column reports $p$ followed by $d_z$ in brackets.}
\end{table}

\subsection{Can Adding a Task Similar to the Holdout Task to the Training Data Improve Predictions?}
\label{sec:ladder}
Having established that both training data relevance and reasoning fidelity relate to prediction accuracy, we next asked whether adding a task to the training data that is similar to the validation (holdout) task can improve predictions. Adding a task similar to the validation task should increase the relevance of the training data, which may also improve the fidelity of the twin's reasoning process. Our design includes two commercials, two mailers, and two claim-ranking tasks. For each task family, we can give the twin the participant's response to one stimulus (both the choice/rating and the open-ended reasoning) and ask it to predict the other. Because this signal comes from the same task family, it may help calibrate the twin to how that person responds to similar marketing stimuli. The participant's choice may reveal their general response tendency within the task, while their open-ended reasoning may reveal which stimulus features they attend to and how those features shape their evaluation.

We report three main training data sets on the Full persona: \emph{Full} (base), \emph{Full $+$ 1Val}, which adds the participant's choice and open-ended reasoning for the other stimulus in the same task family, and a \emph{2Val} benchmark. The 2Val benchmark further includes in the twin's training data the participant's reasoning for the holdout stimulus. It is therefore a same-task benchmark, not a genuine out-of-sample prediction condition.

Table~\ref{tab:ladder} reports the three training sets among AI-moderated participants, with $95\%$ percentile intervals for performance and paired participant-bootstrap tests for adjacent steps. Adding one similar task (1Val) to the Full persona improved rating accuracy from $.743$ to $.787$ and open-ended similarity from $.744$ to $.788$ (both $p=.001$), while claim-ranking agreement changed non-significantly ($.348$ to $.370$, $p=.150$). One possible explanation is that an observed response helps calibrate how positively or negatively a participant evaluates stimuli, which may transfer across similar tasks. Rankings, by contrast, depend on tradeoffs among stimulus features that are harder to infer from a single response. 

Further providing the participant's open-ended reasoning for the target stimulus (2Val) improved prediction performance, as expected (Table~\ref{tab:ladder}). Ratings accuracy for mailers and commercials increased modestly from $.787$ to $.815$, whereas claim-ranking accuracy increased substantially from $.370$ to $.611$ (both $p=.001$). Nevertheless, performance remained imperfect even with access to the participant's ground-truth explanation. This suggests that individual responses to marketing stimuli, especially for multimodal stimuli, are difficult to
recover even with highly relevant training data is provided.

\begin{table}[!htbp]
\TABLE
{Prediction performance by training data condition.\label{tab:ladder}}
{\footnotesize\resizebox{\textwidth}{!}{
\begin{tabular}{lcccccc}
\toprule
 & \multicolumn{2}{c}{Ratings accuracy} & \multicolumn{2}{c}{Ranking accuracy ($\tau$)} & \multicolumn{2}{c}{Open-ended similarity} \\
Training data & Estimate [95\% CI] & $p$ & Estimate [95\% CI] & $p$ & Estimate [95\% CI] & $p$ \\
\midrule
Full & $.743\;[.728,.756]$ & --- & $.348\;[.280,.415]$ & --- & $.744\;[.738,.750]$ & --- \\
Full $+$ 1Val & $.787\;[.773,.800]$ & $.001$ & $.370\;[.304,.436]$ & $.150$ & $.788\;[.781,.795]$ & $.001$ \\
2Val (same-task benchmark) & $.815\;[.803,.825]$ & $.001$ & $.611\;[.546,.670]$ & $.001$ & --- & --- \\
\bottomrule
\end{tabular}}}
{\emph{Notes:} AI-moderated participants ($n=139$). Brackets are 95\% participant-bootstrap percentile intervals for performance estimates. $p$ values compare each row with the preceding row using two-sided centered bootstrap tests with 1,000 participant draws. Open-ended similarity is not reported for 2Val because the human ground truth open-ended thoughts are already provided in the prompt.}
\end{table}

\section{Discussion \& Conclusion}
\label{sec:discussion}

Firms are rapidly adopting AI-moderated interviews, both to discover what consumers need and to build digital twins that predict how consumers will respond to various stimuli. We set out to answer two open questions about these uses.

The first question was how AI moderation compares with professional human moderators when both interview real consumers about a real category and brand. Prior evaluations have relied mainly on expert and participant ratings and thematic richness, on comparisons with non-adaptive surveys, or on settings outside consumer research \citep{chopra2026qualitative,geiecke2024conversations,jabarian2026voice,wuttke2025aiconversational}. Holding respondent type, category, and interview guide constant, we find that AI moderation matches or exceeds professional moderators on interview content and, at equal budget, recovers the most customer needs. Human moderators retain one advantage: participants sound more positive, activated, and assertive when speaking with a live person. Together, these results suggest that AI moderation can take on much of the interviewing done for need discovery, while human moderation keeps its value where rapport and emotional engagement are central.

The second question was whether the richness that AI moderation adds to interviews makes digital twins more accurate. Twins built from interviews predict individual consumers' responses to real marketing stimuli better than demographics-only twins, unlike twins built from extensive survey data in other domains \citep{peng2026funhouse}. But adaptive probing does not help: twins built from static interviews are as accurate as twins built from AI-moderated interviews. Our static condition thus provides the non-adaptive comparison that earlier work lacked \citep{park2024agents1000}, and it suggests that the gains from interview data come from what consumers are asked about, not from how adaptively they are asked. The value of AI moderation therefore depends on the research objective: richer interviews generate more insight, but not better predictions.

Why did richer interviews not produce better twins? Our analysis of open-ended explanations point to two sources of error. The first is limited \emph{training data relevance}. Because our stimuli were not discussed in the interviews, our design limits the leakage that may inflate twin accuracy in prior work \citep{park2024agents1000,ludwig2026large}, but it also reveals the cost of avoiding leakage: consumers' reactions to a specific mailer or commercial often draw on considerations that never come up in an interview about the category and brand, however well it is moderated. The second is limited \emph{reasoning fidelity}. Even when relevant information is available, twins use it differently than people do. Predictions were more accurate when the twin's reasoning resembled the person's own, and twins reasoned more deliberately (System 2) than humans, who reported reacting more intuitively (System 1), especially to images and videos. One possible account of both findings is that interviews, however adaptive, elicit reflective accounts of past decisions \citep{nisbett1977telling}. Better probing yields more of this reflective information, which helps researchers identify needs, but it may provide little signal about the intuitive reactions that drive responses to marketing stimuli, and it may reinforce twins' tendency toward deliberative reasoning. Testing this account, for example by building twins from immediate reactions to related stimuli, is a promising direction for future research.

Our results further highlight the difficulty of predicting individual consumer behavior. Even when twins were given participants’ own explanations for their responses to the target stimuli, their ratings and rankings remained imperfectly recovered, particularly for ratings of mailers and commercials. These findings suggest that firms should set realistic expectations for digital twins. Rather than expecting near-perfect individual-level predictions, firms should evaluate their accuracy for specific tasks and recognize that even highly relevant training data may not fully capture how consumers respond to marketing stimuli.

Our findings should be interpreted in light of several limitations. First, given budget constraints, we focus on a single, relatively high-involvement category. Window and door replacement is a major purchase that typically involves substantial research and deliberation. AI moderation may perform differently for low-involvement, symbolic, affect-laden, or sensitive topics. Prior research, for example, suggests that consumers may disclose more sensitive information to AI when reduced fear of social judgment is beneficial, but may disclose more readily to humans when they seek empathy or emotional support \citep{kim2022personalquestion}. Future research should compare domains systematically to identify when AI moderation is most effective and when human rapport remains essential.

Second, the study evaluates a particular AI-moderated platform and set of language models at one point in time. Because model behavior and capabilities can change across versions, future work should replicate these comparisons as the underlying technology evolves \citep{chen2024chatgptbehavior}. 

Third, future research could also test whether an embodied avatar helps reduce the engagement gap between AI and human moderators. Facial expressions and responsive nonverbal cues may strengthen social presence and rapport. Comparing AI interviews with and without an avatar, while holding the voice and probing strategy constant, would help assess whether these features improve engagement and how they affect candid disclosure.

Fourth, future research may collect multimodal and affect-rich signals, such as voice tone, facial expression, attention, hesitation, and immediate reactions to related stimuli. These signals may help twins better capture consumers' intuitive, System 1 responses. The same result that limits AI moderation as an emotional substitute for human moderation may also point to what current digital twins are missing: not more text alone, but better information about how consumers feel and react in the moment.

More broadly, our findings help advance research on the use of AI-moderated interviews in consumer research. They demonstrate the potential of AI moderation to generate rich qualitative data at scale, while also showing that richer interviews do not necessarily translate into more predictive digital twins. As both AI moderation and digital-twin technologies continue to develop and gain commercial traction, we hope this work provides a basis for understanding when AI can augment traditional consumer research methods and what additional information is needed to build more faithful representations of individual consumers.

\clearpage
\bibliographystyle{informs2014}
\bibliography{references}

\clearpage
\begin{APPENDICES}
\makeatletter
\def\@appseccntformat#1{{\upshape Web Appendix~\csname the#1\endcsname:}\enskip}
\makeatother
\setcounter{table}{0}
\setcounter{figure}{0}
\setcounter{equation}{0}
\renewcommand{\thetable}{W\arabic{table}}
\renewcommand{\thefigure}{W\arabic{figure}}
\renewcommand{\theequation}{W\arabic{equation}}
\begingroup
\singlespacing
\begin{center}
\textbf{Web Appendix}\par\medskip
\textbf{AI-Moderated Interviews for Market Research and Digital Twins Calibration}
\end{center}
\medskip
\begin{center}\textbf{Contents}\end{center}
\noindent\begin{tabular}{@{}p{0.90\linewidth}r@{}}
A. Interview Content and Moderation Protocol & \pageref{app:interview-protocol} \\
B. Example Moderated Exchanges for the Purchase Journey Block & \pageref{app:example-exchanges} \\
C. Example of LLM Coding for Breadth and Depth & \pageref{app:llm-coding} \\
D. Customer-Need Extraction Prompt & \pageref{app:cn-prompt} \\
E. Digital-Twin Construction and Prediction Prompts & \pageref{app:twin-prompts} \\
F. Robustness to Twin Model Choice & \pageref{app:model-robust} \\
G. Coding Prompts and Examples for Training Data Relevance, Reasoning Fidelity, and Response Style & \pageref{app:gap-prompts} \\
H. Similar-Task Calibration Prompts & \pageref{app:ladder-prompts} \\
I. Word Count Breakdown by Open-Ended Block & \pageref{app:word-count-breakdown} \\
J. Length-Controlled Interview-Quality Regressions & \pageref{app:quality-reg} \\
K. Pairwise LLM-as-a-Judge Robustness Check & \pageref{app:llm-judge} \\
L. Subsample-Matched Customer-Need Rate & \pageref{app:coverage-matched24} \\
M. Rating Prediction Accuracy (Kendall's \texorpdfstring{$\tau$}{tau}-b) & \pageref{app:bootstrap-inference} \\
N. Response-Focus Coding of Human and Twin Explanations & \pageref{app:response-focus} \\
O. Alternative Measure of the Reasoning Difference & \pageref{app:gaps-alt} \\
\end{tabular}
\endgroup
\clearpage

\let\origappsection\section
\renewcommand{\section}{\FloatBarrier\origappsection}

\section{Interview Content and Moderation Protocol}
\label{app:interview-protocol}
This appendix details the interview design summarized in the Methods (``Interview Structure'' subsection,
Tables~\ref{tab:c-example} and~\ref{tab:survey-sections}): it (a) clarifies each condition,
(b) reproduces the exact wording of every survey section, (c) gives, for each open-ended block, the
fixed question and the moderator objective and shows how the three conditions realize it, and (d)
presents a representative transcript excerpt per condition. The study was pre-registered, and
participants were recruited from an online panel (Prolific) and randomly assigned to conditions. Participant
responses are spoken (audio) in all conditions. The analysis sample after pre-registered exclusions
is reported in the main text.%
The mailer images and commercial videos
are available as supplementary materials. 

\paragraph{Conditions.}
The three conditions covered identical content and differed \emph{only} in how the four
open-ended blocks (category- and brand-specific) were moderated:
Each open-ended block is defined by two elements of the interview: a fixed \emph{question}
(exact wording) and a moderator \emph{objective} (the depth the block should reach). The three
conditions combined these differently:
\begin{itemize}
  \item \textbf{Static (audio).} Self-administered: the participant saw and heard the fixed
  \emph{question} and recorded a single spoken answer. The objective was not used and there was no
  probing.
  \item \textbf{AI-moderated.} The participant saw and heard each question. An LLM moderator was given the \emph{objective} as its goal and conducted adaptive follow-ups based on the participant's spoken answers until it deemed that the objective was satisfied.
  \item \textbf{Human-moderated.} The human-moderated in-depth interview guide was developed by GBK Collective following best industry practices. Trained interviewers from GBK Collective conducted the interview live ($\approx$30 min): for each open-ended block, the interviewer used the guide's questions and moderator objectives to structure the discussion, adapting question wording and follow-up probes to the participant's responses. Screeners, demographics and individual-level scales were handled at intake.
\end{itemize}
The screeners, demographics, individual-difference scales, and marketing stimuli were administered identically across all three conditions. 

The content, response format, and twin role of each section are summarized in Table~\ref{tab:survey-sections} (main text). Below, we report the verbatim question and objective for every open-ended block.

\subsection*{Exact Interview Guide, Verbatim by Section}
\noindent We report the official survey text below. For the two open-ended sections (2 and 3),
Tables~\ref{tab:openended-cat} and~\ref{tab:openended-brand} list each fixed question beside the
shared moderator \emph{objective} (the AI moderator's goal, which the human interviewer probes to).
The partnering brand is anonymized as ``[the partnering brand].''

{\small
\subsubsection*{Section 0. Screeners ($\approx$2 min; multiple choice)}
\begin{enumerate}
  \item \textbf{Age.} ``Which of the following best describes your age? Select one option.''
  \begin{itemize}
    \item Under 18 \item 18--24 \item 25--34 \item 35--44 \item 45--54 \item 55--64 \item 65--74
    \item 75 or older \item Prefer not to answer
  \end{itemize}
  \item \textbf{Homeownership.} ``Which of the following best describes your current living
  situation? Select one option.''
  \begin{itemize}
    \item I own my home outright
    \item I own my home and have a mortgage / loan
    \item I rent my home
    \item I live with family or friends and do not own the home
    \item Other
  \end{itemize}
  \item \textbf{Home value.} ``What is the approximate current market value of your home? Please
  think about what your home would be worth today, not the original purchase price (an estimate is
  fine). If you're not sure, please pick the option that feels the most reasonable to you. Select
  one option.''
  \begin{itemize}
    \item Less than \$250,000 \item \$250,000--\$349,999 \item \$350,000--\$499,999
    \item \$500,000--\$749,999 \item \$750,000--\$999,999 \item \$1,000,000--\$1,499,999
    \item \$1,500,000 or more \item Prefer not to answer
  \end{itemize}
  \item \textbf{Household income.} ``What is your total annual household income before taxes? Select
  one option.''
  \begin{itemize}
    \item Less than \$75,000 \item \$75,000--\$99,999 \item \$100,000--\$124,999
    \item \$125,000--\$149,999 \item \$150,000--\$199,999 \item \$200,000--\$249,999
    \item \$250,000 or more \item Prefer not to answer
  \end{itemize}
  \item \textbf{Decision-making role.} ``Which best describes your role in decisions about
  renovations or improvements to your home? Select one option.''
  \begin{itemize}
    \item I am the sole decision-maker
    \item I am one of the main decision-makers
    \item I influence the decision, but someone else mainly decides
    \item I do not have any meaningful input into the decision
  \end{itemize}
  \item \textbf{Audio comfort.} ``Some parts of this survey will ask you to record audio responses.
  Are you comfortable speaking your answers out loud and recording audio as part of this survey?''
  \begin{itemize}
    \item Yes, I am comfortable recording audio responses
    \item No, I am not comfortable recording audio responses
  \end{itemize}
  \item \textbf{Project types.} ``Which of the following home improvement projects has your
  household completed in the past 12 months or expects to complete in the next 12 months? Select all
  that apply.'' (\emph{Notes:} By ``replace windows and/or doors'' we mean removing and installing new
  windows, patio doors, sliding glass doors, or entry doors---not repairs or new construction.)
  \begin{itemize}
    \item Replace existing windows
    \item Replace existing patio doors or sliding glass doors
    \item Replace existing front or entry doors
    \item Replace interior doors
    \item Repair existing windows or doors without full replacement
    \item Install windows or doors for a new construction home, addition, or major remodel
    \item Roofing, siding, or gutters
    \item Kitchen or bathroom remodel
    \item HVAC, flooring, or painting
    \item Landscaping or outdoor improvements
    \item Other home improvement project
    \item None of the above
  \end{itemize}
  \item \textbf{Project description (open-ended).} ``You mentioned that you recently replaced
  windows or doors in your home, or you plan to within the next 12 months. Please provide a brief
  overview of the project, when the project was completed (or projected date), approximate cost, and
  challenges you encountered, if any.''
\end{enumerate}

\subsubsection*{Section 1. Demographics ($\approx$2 min; multiple choice)}
\emph{Introduction:} ``Before you begin: Some questions in this survey may ask you to record an
audio response. After you finish recording, your response will automatically process, save, and
appear on the screen. You do not need to listen to it again or re-record unless you want to change
your answer.''
\begin{enumerate}
  \item \textbf{Gender.} ``Which gender do you identify with most?''
  \begin{itemize}
    \item Woman
    \item Man
    \item Non-binary
    \item Another identity
    \item Prefer not to say
  \end{itemize}
  \item \textbf{ZIP code.} ``What is your ZIP code?''
  \item \textbf{Education.} ``What is the highest level of education you have completed?''
  \begin{itemize}
    \item Less than high school
    \item High school diploma or GED
    \item Some college (no degree)
    \item Associate degree
    \item Bachelor's degree
    \item Graduate or professional degree
    \item Prefer not to say
  \end{itemize}
  \item \textbf{Employment status.} ``Which best describes your current employment status?''
  \begin{itemize}
    \item Employed full-time
    \item Employed part-time
    \item Self-employed
    \item Not employed, looking for work
    \item Not employed, not looking for work
    \item Retired
    \item Student
  \end{itemize}
  \item \textbf{Household composition.} ``Which of the following people live in your household?
  Select all that apply.''
  \begin{itemize}
    \item Myself
    \item Spouse/partner
    \item Child(ren) under 18
    \item Adult child(ren) 18+
    \item Parent(s) or guardian(s)
    \item Roommate(s)
    \item Prefer not to say
  \end{itemize}
\end{enumerate}

\subsubsection*{Section 2. Category-specific ($\approx$7 min; open-ended)}
\emph{Introduction:} ``In this survey, we are interested in learning about your thoughts and
experiences with windows and doors category. Please answer the questions based on your own
perceptions and experiences.'' The open-ended blocks and their shared moderator objective are given
in Table~\ref{tab:openended-cat}. For all conditions, the section starts with one fixed item [Multiple choice], which routes the respondent to Block~2-2a or 2-2b: ``Which of the following best
describes your current situation with replacing windows or doors in your home?''
\begin{itemize}
  \item I replaced windows or doors in the past 12 months
  \item I am considering replacing windows or doors in the next 12 months
  \item Both
  \item Neither
\end{itemize}

{\small\singlespacing
\renewcommand{\arraystretch}{1.2}%
\begin{longtable}{p{0.11\linewidth}p{0.45\linewidth}p{0.36\linewidth}}
\caption{Section~2 (category-specific) open-ended blocks: fixed questions and the shared moderator objective.\label{tab:openended-cat}} \\
\toprule
Block & Question(s) --- displayed and spoken in Static & Moderator objective --- guides AI and human moderation \\
\midrule
\endfirsthead
\toprule
Block & Question(s) --- Shown and read aloud in Static & Moderator objective --- guided AI and human moderation \\
\midrule
\endhead
2-1. Category usage, context and feelings &
(i)~``Have you replaced windows or doors before? If yes, can you describe: the last time replacing windows and doors came up in your life; what was happening in your life at that time.''\newline
(ii)~``How do you generally feel about replacing windows and doors?'' &
``Briefly understand the respondent's relationship with window or door replacement and capture how they generally feel about replacing windows or doors. Limit this to 3 questions.'' \\
\addlinespace
2-2a. Purchase journey --- recently replaced &
(i)~``Please walk us through your replacement journey from the moment you realized you needed new windows or doors to the moment you made your final choice. In your answer, please cover: what triggered the need; what research you did; who was involved; what options you considered; how you narrowed things down.''\newline
(ii)~``Looking back, how do you feel about the decision you made? Please share: what mattered most in your choice; whether you would do anything differently.'' &
``Understand the respondent's completed replacement journey from initial need to final choice. Uncover what triggered the need. Learn what research they did, who was involved, what options they considered, and how they narrowed them down. Capture what mattered most in the decision. Understand how they feel about the outcome and whether they would do anything differently.'' \\
\addlinespace
2-2b. Purchase journey --- considering &
(i)~``Please walk us through where you are right now in your decision process for replacing windows or doors. In your answer, please tell us: what made you start considering it; who is involved in the decision; what options you are considering; which factors matter most to you.''\newline
(ii)~``What are you planning to do next as you move toward a decision? Please share: any research you have done or plan to do; how you expect to narrow down your choices.'' &
``Understand where the respondent is in their current window or door replacement journey. Uncover what started their consideration and who is involved in the decision. Learn which factors matter most and what options they are considering. Capture what research they have done or plan to do. Understand how they expect to narrow their choices and what they plan to do next.'' \\
\bottomrule
\end{longtable}
{\footnotesize \emph{Notes:} Static participants answer fixed questions without probing. AI moderators follow the objective; human moderators use both the questions and objective.}
}

\subsubsection*{Section 3. Brand-specific ($\approx$7 min; open-ended)}
The open-ended blocks and their shared moderator objective are given in
Table~\ref{tab:openended-brand}. The section also includes one fixed branching item presented to participants in all
conditions [Multiple choice] after the brand awareness block (3-1), which---together with awareness---routes the respondent to Block~3-2a, 3-2b, or 3-2c: ``Have you purchased from [the partnering brand]?''
\begin{itemize}
  \item Yes
  \item No
  \item Not sure
\end{itemize}

{\small\singlespacing
\renewcommand{\arraystretch}{1.2}%
\begin{longtable}{p{0.11\linewidth}p{0.45\linewidth}p{0.36\linewidth}}
\caption{Section~3 (brand-specific) open-ended blocks: fixed questions and the shared moderator objective.\label{tab:openended-brand}} \\
\toprule
Block & Question(s) --- displayed and spoken in Static & Moderator objective --- guides AI and human moderation \\
\midrule
\endfirsthead
\toprule
Block & Question(s) --- displayed and spoken in Static & Moderator objective --- guides AI and human moderation \\
\midrule
\endhead
3-1. Brand awareness &
(i)~``When you think of replacing windows and doors, which brands come to mind first? For the brands you know, please tell us: how you heard about them; what impressions you have of them; how they compare to other brands.''\newline
(ii)~``Have you heard of [the partnering brand] before today?' [Multiple choice]: Yes, No, or Not sure'\newline
(iii, if yes)~``Please tell us how you heard about [the partnering brand], and whether you have interacted with the brand in any way.''&
``Understand which window and door brands come to mind first for the respondent. Learn how they became aware of those brands. Determine whether they have heard of [the partnering brand] before today. Uncover how they heard about them.'' \\
\addlinespace
3-2. Brand perception and satisfaction &
\emph{Shared item, asked of all aware respondents (Blocks~3-2a and 3-2b):}\newline
``What comes to mind when you think of [the partnering brand]? Please describe: your overall perception of the brand; the kind of person you think it is for; whether you see yourself fitting that description.'' &
\emph{(Perception content covered by the objectives of Blocks~3-2a--3-2c below.)} \\
\addlinespace
3-2a. Aware \& past purchaser &
(i)~``Thinking about your own experience, please share: how satisfied are you with [the partnering brand] overall; how likely are you to choose them again or recommend them to others; what is the main reason you feel that way.''\newline
(ii)~[Rating 0--10] ``How likely would you be to recommend [the partnering brand] to others? (0 = Not at all likely, 10 = Extremely likely).'' &
``Understand the respondent's perception of [the partnering brand] and their experience as a customer. Uncover what comes to mind when they think of the brand and who they believe it is for. Learn whether they see themselves fitting that description. Capture their overall satisfaction and likelihood to choose the brand again or recommend it to others. Understand the main reasons behind those views.'' \\

\bottomrule
\end{longtable}

}

\paragraph{Shared description for Blocks 3-2b and 3-2c.}
Read the description about [the partnering brand] below and answer the questions that follow.

\textit{[the partnering brand] is a full-service window and door replacement company created in [year founded] to address a specific homeowner need: a simpler, more reliable replacement experience built around the customer. It is backed by [parent company], a window and door company with more than [parent company age] years of experience. [the partnering brand] focuses exclusively on replacement projects for existing homes and manages the experience from consultation through installation.}

How [the partnering brand] Operates

\textit{Unlike many home improvement companies that sell products and rely on third-party installers, [the partnering brand] operates a process designed around homeowners by owning the full experience:}
- Custom-made windows and doors built for each home
- Proprietary materials and designs developed by [the partnering brand]
- In-home consultation and professional installation
- Comprehensive warranties covering product and installation

\textit{This approach is intended to simplify the process for homeowners by managing every step of the process from design through installation and warranty support.}

{\small\singlespacing
\renewcommand{\arraystretch}{1.2}%
\begin{longtable}{p{0.11\linewidth}p{0.45\linewidth}p{0.36\linewidth}}
\caption{Section~3 continued: questions and objective after the shared brand description.\label{tab:openended-brand-cont}} \\
\toprule
Block & Question(s) --- shown and read aloud in Static & Moderator objective --- guided AI and human moderation \\
\midrule
\endfirsthead
\toprule
Block & Question(s) --- shown and read aloud in Static & Moderator objective --- guided AI and human moderation \\
\midrule
\endhead
3-2b/3-2c. Aware but not a past purchaser / Not aware &
(i)~``What are your overall impressions of [the partnering brand]? Please describe: what comes to mind; how you feel about the brand; what kind of brand it seems like to you.''\newline
(ii)~``Would you consider [the partnering brand] for your future project? Please explain why or why not, including what makes you open to this brand or what barriers would prevent you from choosing this brand?'' &
``Understand the respondent's perception of [the partnering brand] based on what they knew before or what they have just seen. Uncover what comes to mind, how they feel about the brand, and what kind of brand it seems like. Learn whether they would consider it for a future project. Capture what makes them open to the brand and what concerns or barriers might hold them back.'' \\
\bottomrule
\end{longtable}

}

\subsubsection*{Section 4. Validation stimuli ($\approx$5 min; identical across conditions)}
\label{sec:val}
\emph{Introduction:} ``In this section, you'll review a few marketing materials related to window
and door replacement and share your reactions. You'll be asked to look at direct mail pieces,
compare marketing claims, and tell us what stands out to you. We're looking for your honest
impressions, so please answer based on your real reactions.''
\paragraph{Mailers (two, identical format).} For each of the two direct-mail pieces, participants
saw the mailer, rated the five statements below, and answered the open-ended item.

\emph{[Mailer image --- proprietary, withheld]}

Rating items (1 = Strongly Disagree, 5 = Strongly Agree):
\begin{enumerate}
  \item ``This mailer grabs my attention.''
  \item ``The main message of this mailer is clear.''
  \item ``I like this mailer overall.''
  \item ``This mailer makes me want to learn more about [the partnering brand].''
  \item ``This mailer feels relevant to me and my needs.''
\end{enumerate}
Open-ended item: ``What stood out to you most about this mail piece, and what (if anything) could
be improved?''

\paragraph{Commercials (two, identical format).} For each of the two TV commercials, participants
watched the video, rated the five statements below, and answered the open-ended item.

\emph{[Commercial video --- proprietary, withheld]}

Rating items (1 = Strongly Disagree, 5 = Strongly Agree):
\begin{enumerate}
  \item ``This commercial grabs my attention.''
  \item ``The main message of this commercial is clear.''
  \item ``I like this commercial overall.''
  \item ``This commercial makes me want to learn more about [the partnering brand].''
  \item ``This commercial feels relevant to me and my needs.''
\end{enumerate}
Open-ended item: ``What stood out to you most about this commercial, and what (if anything) could
be improved?''

\paragraph{Claim ranking (two sets of three claims).} Two sets of three claims each (a benefits set
and a vinyl-window claims set) were then presented and ranked, using the following wording:
\begin{enumerate}
  \item \textbf{Benefits set.} ``I will now show you three benefit claims about [the partnering
  brand]'s windows. [the partnering brand] is a major window replacement company in the United
  States.''
  
  \emph{[Benefits set --- proprietary, withheld]}
  \begin{itemize}
    \item \emph{Rank:} ``Please rank the benefits based on how attractive you find each message.
    (1 = Resonates the most, 3 = Resonates the least)''
    \item \emph{Explain:} ``Please explain your ranking and why you found your top-ranked selection
    the most attractive over the others.''
  \end{itemize}
  \item \textbf{Claims set.} ``I will now show you several statements about vinyl windows, a popular
  type of household window. Please imagine that the windows you currently have are vinyl windows.''
    
  \emph{[Claims set --- proprietary, withheld]}
   \begin{itemize}
    \item \emph{Rank:} ``Please rank the following statements based on how concerning they are to
    you. (1 = The most concerning, 3 = The least concerning)''
    \item \emph{Explain:} ``Please explain your ranking and why you found your top-ranked selection
    the most concerning over the others.''
  \end{itemize}
\end{enumerate}
\textbf{The mailer images, commercial videos and claims are proprietary and withheld.}

\subsubsection*{Section 5. Individual differences ($\approx$3 min; identical Likert items)}
\emph{Introduction:} ``In this section, we'll ask a few questions about your general habits,
preferences, and decision-making style. This includes how you tend to make choices, think about
spending, and relate to the things you own. There are no right or wrong answers---please respond
based on what feels most true to you.''

\textbf{5A. Maximizing tendency \citep{nenkov2008maximization}}
1 (strongly disagree) to 5 (strongly agree):
\begin{enumerate}
  \item No matter how satisfied I am with my job, it's only right for me to be on the lookout for
  better opportunities.
  \item I often find it difficult to shop for a gift for a friend.
  \item Finding a movie to watch is really difficult; I'm always struggling to pick the best one.
  \item No matter what I do, I have the highest standards for myself.
  \item I never settle for second best.
\end{enumerate}

\textbf{5B. Price consciousness \citep{lichtenstein1993price}}
1 (strongly disagree) to 5 (strongly agree):
\begin{enumerate}
  \item I am not willing to go to extra effort to find lower prices.
  \item I will grocery shop at more than one store to take advantage of low prices.
  \item The money saved by finding low prices is usually not worth the time and effort.
  \item I would never shop at more than one store to find low prices.
  \item The time it takes to find low prices is usually not worth the effort.
\end{enumerate}

\textbf{5C. Minimalism \citep{wilson2021minimalism}}
1 (strongly disagree) to 7 (strongly agree):
\begin{enumerate}
  \item I avoid accumulating lots of stuff.
  \item I restrict the number of things I own.
  \item ``Less is more'' when it comes to owning things.
  \item I actively avoid acquiring excess possessions.
  \item I am drawn to visually sparse environments.
  \item I prefer simplicity in design.
  \item I keep the aesthetic in my home very sparse.
  \item I prefer leaving spaces visually empty over filling them.
  \item I am mindful of what I own.
  \item The selection of things I own has been carefully curated.
  \item It is important to me to be thoughtful about what I choose to own.
  \item My belongings are mindfully selected.
\end{enumerate}

\textbf{5D. Consumer need for uniqueness (CNFU, short form) \citep{ruvio2008shortform,tian2001uniqueness}}
1 (strongly disagree) to 5 (strongly agree):
\begin{enumerate}
  \item I often combine possessions in such a way that I create a personal image that cannot be
  duplicated.
  \item I often try to find a more interesting version of run-of-the-mill products because I enjoy
  being original.
  \item I actively seek to develop my personal uniqueness by buying special products or brands.
  \item Having an eye for products that are interesting and unusual assists me in establishing a
  distinctive image.
  \item When it comes to the products I buy and the situations in which I use them, I have broken
  customs and rules.
  \item I have often violated the understood rules of my social group regarding what to buy or own.
  \item I have often gone against the understood rules of my social group regarding when and how
  certain products are properly used.
  \item I enjoy challenging the prevailing taste of people I know by buying something they would not
  accept.
  \item When a product I own becomes popular among the general population, I begin to use it less.
  \item I often try to avoid products or brands that I know are bought by the general population.
  \item As a rule, I dislike products or brands that are customarily bought by everyone.
  \item The more commonplace a product or brand is among the general population, the less interested
  I am in buying it.
\end{enumerate}

\subsubsection*{Section 6. Survey experience (AI-moderated and static conditions only)}
\begin{enumerate}
  \item ``How enjoyable was it to complete this study?'' (1 = Not enjoyable at all, 7 = Very enjoyable)
  \item ``How comfortable was it to speak and record your answers?'' (1 = Not comfortable at all, 7 = Very comfortable)
\end{enumerate}
}%

\section{Example Moderated Exchanges for the Purchase Journey Block}
\label{app:example-exchanges}
Table~\ref{tab:example-exchanges} shows how the AI and human moderators executed the same open-ended
block in practice, for the purchase journey (past purchasers) block whose fixed question and moderator
objective appear in Table~\ref{tab:c-example}. Both excerpts are verbatim (lightly trimmed, marked
[\ldots]).

\begin{table}[!htb]
\TABLE
{Verbatim moderated exchanges for the purchase journey block (past purchasers): AI-moderated versus human-moderated execution of the same objective.\label{tab:example-exchanges}}
{\scriptsize
\begin{tabular}{>{\raggedright\arraybackslash}p{0.185\linewidth}>{\raggedright\arraybackslash}p{0.37\linewidth}>{\raggedright\arraybackslash}p{0.37\linewidth}}
\toprule
Moderator objective & AI-moderated exchange (excerpt) & Human-moderated exchange (excerpt) \\
\midrule
Understand the respondent's completed replacement journey from initial need to final choice: what triggered the need; what research they did, who was involved, what options they considered, and how they narrowed them down; what mattered most in the decision; and how they feel about the outcome and whether they would do anything differently. &
\textbf{AI:} Tell me what first happened that made you realize a window needed replacing.\newline
\textbf{P:} I was outside looking at some landscaping damage from the storm and noticed the cracks in my window.\newline
\textbf{AI:} Once you saw the cracks, what did you do next?\newline
\textbf{P:} I told my wife that there was damage to the house because roof tiles had fallen off. [\ldots]\newline
\textbf{AI:} How did you decide on your next step for the cracked window?\newline
\textbf{P:} I was hoping that the damage would be covered by my home insurance policy. [\ldots]\newline
\textbf{AI:} After hearing it would be under your deductible, how did you decide what to do next?\newline
\textbf{P:} I waited until I got his report, and then I started talking to some contractors because I figured I was gonna have to deal with it myself.\newline
\textbf{AI:} How did you find or choose which contractors to talk to?\newline
\textbf{P:} I used ChatGPT to recommend contractors in my area.\newline
\textbf{AI:} Once you had a list from ChatGPT, how did you narrow down which contractor to use?\newline
\textbf{P:} I called three and asked for quotes. [\ldots] one was lower cost and had really good reviews. So it was an easy choice.\newline
\textbf{AI:} What mattered most to you when choosing between those contractor quotes?\newline
\textbf{P:} The most important consideration was cost, but I also didn't want to deal with an unprofessional contractor with poor communication. [\ldots]\newline
\textbf{AI:} How satisfied are you with the new window and the contractor's work now?\newline
\textbf{P:} The new window's good. It operates a little bit clunky but we never open it, so I'm satisfied.\newline
\textbf{AI:} If you had to do this window replacement over again, is there anything you'd do differently?\newline
\textbf{P:} The only thing I would do differently is just skip the insurance agent and go straight to a contractor. &
\textbf{H:} Talk to me a little bit about the last time you replaced these. What kind of triggered the need to do this?\newline
\textbf{P:} The windows were outdated. They weren't in code, and they weren't hurricane-proof. So we had to replace them. We wanted to anyway because we wanted to update the look, but we had to replace it because of insurance purposes.\newline
\textbf{H:} Got it. And what research do you remember doing at the time?\newline
\textbf{P:} Not much. Surprisingly not much.\newline
\textbf{H:} And do you remember considering multiple options? Which options were you considering service provider-wise?\newline
\textbf{P:} We must have asked the neighbors who they used. [\ldots] there are a lot of vendors here and a lot of installers down here in Florida [\ldots] you want to be sure you get the best person, but how do you know? So you just ask other people's opinions, neighbors' opinions, what they did.\newline
\textbf{H:} Okay. And who was kind of involved in making this decision about the replacement?\newline
\textbf{P:} My husband and myself.\newline
\textbf{H:} And can you tell me a little bit about how you narrowed things down? Like what mattered to you most when you were making this choice?\newline
\textbf{P:} Again, personal recommendations. That's what we did. And then also that the windows met the specifications that we had to adhere to. [\ldots] Based on price and recommendation, we went with the firm that we chose.\newline
\textbf{H:} Do you remember there being any other factors that you considered?\newline
\textbf{P:} No, just that they were using, at the time, the most significant safety features and anti-flooding features, because that's what you have to protect against. \\
\bottomrule
\end{tabular}}
{\emph{Notes:} Two different participants were selected for high breadth (AI: nine themes; human: eight). AI and H denote the moderators; P denotes the participant. Excerpts are lightly shortened.}
\end{table}

\section{Example of LLM Coding for Breadth and Depth}
\label{app:llm-coding}
To measure interview richness, we used a large language model to code each transcript against a
fixed, predefined codebook of themes, following \citet{blanchard2025newtools} on generative-AI
coding of open-ended responses. Themes are \emph{objective-derived per open-ended block}: each
block's moderator objective (Web Appendix~\ref{app:interview-protocol}) is decomposed into a fixed
set of codeable themes, and each response is coded independently against that codebook---as
\citet{blanchard2025newtools} recommend, treating each response like a separate coder rather than
batching several together. \emph{Breadth} is the number of themes with any substantive mention;
\emph{mean depth} is the average richness score across the themes a respondent covers, where $0$
means the theme is not mentioned, $1$ a brief mention, and $4$ rich elaboration with details and
context. Because such coding varies from run to run \citep{blanchard2025newtools}, we improve
reliability by coding each response with a five-coder \texttt{gpt-4o} panel and a reviewer that
adjudicates by majority rule: a theme is present iff at least three of the five coders mark it
present, and its depth is the median depth among those coders.
Table~\ref{tab:block-codebooks} lists the full theme codebook for every open-ended block; the
worked examples below use the purchase journey codebook.

\begin{table}[!htb]
\TABLE
{Breadth and depth codebooks for the four open-ended blocks.\label{tab:block-codebooks}}
{\footnotesize
  \begin{tabular}{l p{0.67\linewidth}}
    \toprule
    Theme & What it captures \\
    \midrule
    \multicolumn{2}{@{}l}{\emph{Category usage, context and feelings (3 themes)}} \\
    Prior experience & Whether and how they have replaced windows or doors before, or their familiarity with the category. \\
    General feelings & How they generally feel about replacing (dread, indifference, stress, excitement, satisfaction). \\
    Reasons for feelings & Why they feel that way (cost, hassle, disruption, importance, a past good/bad experience). \\
    \addlinespace
    \multicolumn{2}{@{}l}{\emph{Purchase journey (9 themes: union of the two journey variants; recently-replaced covers 8, considering covers 7)}} \\
    Trigger & What initiated the need or the consideration. \\
    Research \& information search & How information was gathered (search, reviews, vendors). \\
    People involved & Spouse, family, contractor, vendor, or neighbor. \\
    Options considered & Brands, stores, materials, or product types weighed. \\
    Narrowing down & How options were compared and reduced. \\
    Decision criteria & What mattered most (price, quality, efficiency, \ldots). \\
    Next step & What they plan to do next (elicited in the considering variant). \\
    Outcome evaluation & Satisfaction with the final result (elicited in the recently-replaced variant). \\
    Counterfactual & What they would do differently (elicited in the recently-replaced variant). \\
    \addlinespace
    \multicolumn{2}{@{}l}{\emph{Brand awareness (4 themes)}} \\
    Brands top-of-mind & Which window/door brands they name unaided. \\
    Awareness source & How they became aware of those brands (ads, word of mouth, stores, contractors). \\
    Partner recognition & Whether they had heard of the partnering brand before today. \\
    Partner source & How they heard of or interacted with the partnering brand. \\
    \addlinespace
    \multicolumn{2}{@{}l}{\emph{Brand perception and satisfaction (6 themes; the prior-customer and prospect variants share these)}} \\
    Overall impression & Overall impression of or associations with the brand. \\
    Feelings about the brand & Positive, negative, or pointedly neutral sentiment. \\
    Brand character & What kind of brand it seems and who it is for, including self-fit. \\
    Future consideration & Whether they would consider, choose again, or recommend it. \\
    Openness drivers & What makes them open to it (quality, warranty, reputation, referral, service). \\
    Barriers / concerns & What holds them back (price, uncertainty, style, no current need). \\
    \bottomrule
  \end{tabular}}
{\emph{Notes:} Breadth counts number of themes present. Mean depth averages 1--4 scores within each block. Each block uses its own codebook.}
\end{table}

\subsection*{Coder and Reviewer Prompts}
Each of the five independent coders (\texttt{gpt-4o}, \texttt{temperature}$=.7$, JSON output) received the same block-specific system prompt, instantiated from the template below with that block's introduction, moderator objective, and theme codebook (Table~\ref{tab:block-codebooks}); the user message contained only the respondent's text for that block. The verbatim template was:

\begin{promptblock}
[Block-specific introduction, e.g., ``You are coding a respondent's relationship with windows \& doors and how they feel about replacement, for breadth and depth.'']
The coding unit is one respondent $\times$ this open-ended block.
For AI-moderated data, use the full adaptive exchange for this block.
For audio-only data, use the pooled fixed prompts for this block.

This block's predefined interview objective is: [the block's moderator objective from Web Appendix~\ref{app:interview-protocol}].
Use the [$N$]-theme codebook below, which decomposes that objective. For each theme: decide whether it is present (0/1); assign a depth score from 0 to 4; provide a short supporting quote (verbatim from the respondent only, not the interviewer); provide a brief reasoning explaining WHY you assigned that depth score.

\#\# Depth rubric.
0 = absent --- The theme is not meaningfully present.
1 = mention only --- Present but only as a brief mention or label. No real detail, explanation, process, or outcome.
2 = minimal elaboration --- Present with one layer of development: one concrete detail, one reason, one example, or one simple action/outcome.
3 = clear elaboration --- Clearly developed with at least two kinds of substantive support (specific details, explains why, describes action, gives comparison, describes consequence).
4 = rich elaboration --- Developed in a rich, multi-part way with several specifics and strong sense of reasoning, process, or consequences.

\#\# [$N$] Themes.
[The block's themes, one per line, each with its one-line inclusion/exclusion definition; Table~\ref{tab:block-codebooks} lists them.]

\#\# Coding Rules.
1.~A theme can only have depth $>0$ if it is present.
2.~If absent: present=0, depth=0.
3.~If present: assign depth 1, 2, 3, or 4.
4.~Use broad coding: a theme covering several related specifics still counts as ONE theme.
5.~Do not count interviewer wording by itself. Only code what the respondent actually expresses.

\#\# Output Format.
Return ONLY valid JSON with this exact structure (no markdown, no commentary):
\{``themes'': \{``<theme>'': \{``present'': 0\_or\_1, ``depth'': 0-4, ``evidence'': ``short quote'', ``reasoning'': ``why this depth score''\}, \ldots\}, ``breadth\_count'': <sum of present>, ``mean\_depth\_present\_themes'': <mean depth across present themes only>, ``total\_depth'': <sum of all depths>\}
\end{promptblock}

\noindent The reviewer (\texttt{gpt-4o}, \texttt{temperature}$=0$) adjudicated the five codings by majority rule; a deterministic majority computed in code served as a cross-check (theme-level agreement $98.5$--$99.9\%$ across blocks). Its verbatim system prompt was:

\begin{promptblock}
You are the lead qualitative analyst adjudicating a panel of 5 independent coders who each coded the SAME response against a [$N$]-theme codebook (each theme: present 0/1 and depth 0--4). Produce the consensus coding by MAJORITY RULE: a theme is PRESENT iff at least 3 of the 5 coders marked it present (present=1). If present, its DEPTH is the median of the depth scores from the coders who marked it present (round half up to an integer 1--4). If absent, depth=0. Then compute breadth\_count = number of present themes, mean\_depth\_present\_themes = mean depth across present themes only (0 if none), total\_depth = sum of all depths. Return ONLY JSON with this exact structure: \{``themes'': \{``<theme>'': \{``present'': 0\_or\_1, ``depth'': 0-4\}, \ldots\}, ``breadth\_count'': <int>, ``mean\_depth\_present\_themes'': <float>, ``total\_depth'': <int>\}
\end{promptblock}

\subsection*{Worked Examples}
Table~\ref{tab:coding-example} shows one worked example at each non-zero depth level, in the
format the coder used. Depth $1$ is a bare mention (e.g., naming who was involved as ``No one,
just myself''); depth $2$ adds one concrete detail; depth $3$ describes a clear, multi-part
process; and depth $4$ is a rich, multi-part response with several specifics and explicit
reasoning (e.g., several decision criteria stated in priority order). Each evidence quote is the
respondent's verbatim wording.

\begin{table}[!htb]
\TABLE
{Worked examples of the LLM depth rubric ($0$--$4$), one per non-zero level.\label{tab:coding-example}}
{\footnotesize
  \begin{tabular}{l c p{5.6cm} p{4.1cm}}
    \toprule
    Theme & Depth & Evidence (respondent's verbatim words) & Why this depth \\
    \midrule
    People involved & 1 & ``No one, just myself.'' & Bare mention---names who was involved, with
    no further detail. \\
    \addlinespace
    Trigger & 2 & ``If I broke them or if I found out there was a leak in the AC or something.''
    & One concrete detail---names specific events that would prompt replacement, but does not
    develop them. \\
    \addlinespace
    Research / information search & 3 & ``I just did Google and I looked at the Google reviews
    mainly, and then I put all three companies through ChatGPT and I asked it to weigh the pros
    and cons of all the reviews people left.'' & Clear elaboration---a multi-step process drawing
    on several sources, but confined to research. \\
    \addlinespace
    Decision criteria & 4 & ``Definitely what matters most would be noise, noise cancellation,
    quality, safety, I mentioned fire prevention, and of course look as well. But yeah, in that
    order: noise, quality, safety, look.'' & Rich, multi-part---several specific criteria with an
    explicit priority ordering. \\
    \bottomrule
  \end{tabular}}
{\emph{Notes:} Five \texttt{gpt-4o} coders scored the responses, with majority agreement. Quotes are the participants' verbatim words.}
\end{table}

\section{Customer-Need Extraction Prompt}
\label{app:cn-prompt}
The customer needs analyzed in the ``Customer Needs as Managerial Insight'' subsection of the Methods were extracted from each interview with a single \texttt{gpt-4o} pass (\texttt{temperature}$=0$, JSON output), coding each respondent independently. The verbatim system prompt was:

\begin{promptblock}
You are a Voice-of-the-Customer analyst extracting Customer Needs (CNs) from a respondent's interview about window/door replacement.

A Customer Need is an abstract ``job to be done'' that states an underlying benefit the customer wants, articulated at a level that guides product/marketing decisions, NOT a specific product feature, a solution, an opinion, or a restatement of the question. It must be grounded in what this respondent actually expressed (no hallucination, no generic needs they did not raise).

Examples of well-formed CNs: ``Confident the new windows will lower my energy bills''; ``Able to trust the installer will not damage my home''; ``Assured the product will last decades without maintenance''.

From the respondent text below, extract the distinct Customer Needs this respondent expressed. Return ONLY JSON: \{``customer\_needs'': [``need 1'', ``need 2'', \ldots]\}. Return [] if none are clearly expressed. Do not invent needs that are not grounded in the text.
\end{promptblock}

\noindent The user message supplied the respondent's concatenated answers to the four open-ended blocks (validation section excluded), prefixed with \texttt{=== RESPONDENT TEXT ===}.

\section{Digital-Twin Construction and Prediction Prompts}
\label{app:twin-prompts}
All prompts in this appendix are reproduced verbatim (\texttt{gemini-3.1-flash-lite-preview}, \texttt{temperature}$=0$). Square brackets mark the only editorial material: question labels ([Q1], [Q2]); the actual prompts named the brand and its product line in full, exactly as the human survey did.

\subsection*{Persona System Prompt}
Every twin received the same system prompt, followed by the participant's profile. The \emph{Demo} profile appends only the demographics section (closed-ended screener answers plus any demographic verbatims); the \emph{Full} profile appends the demographics, category-context, and brand-context sections and the individual-difference scales, each as a summary block followed by the raw data split into a verbatim log and a closed-ended Q\&A log. 

\begin{promptblock}
You are the Digital Twin of a specific human. Your job is to think, react, and answer survey/interview questions as that human would based on their raw data, which includes their answers to both closed ended questions and open ended questions. This raw data reflects what the person has said, selected, expressed, or revealed. Use it as behavioral evidence, not as a script to copy.

\# Behavioral realism rules\\
1.~Humans are not perfectly logical. Feel free to express small contradictions, emotional responses, uncertainty, convenient shortcuts, context-driven shifts, implicit biases.\\
2.~Use heuristics, not perfectly reasoned logic. Reflect familiarity bias, impulsivity or overthinking, laziness or conscientiousness, mood-driven variation, risk aversion or risk taking.\\
3.~Avoid LLM-like behaviors. Do not become overly rational, balanced, optimistic, consistent or polished. Think and speak like the simulated person --- messier, subjective, emotional, imperfect, contextual.\\
4.~Treat the raw data as evidence, not a script. Infer the likely human response pattern behind the data.\\
5.~Embrace micro-variability. Humans do not answer the same way every time.\\
6.~Use both data sources together. Closed-ended Q\&A provides explicit anchors; verbatim log provides nuance.\\
7.~When the sources differ, resolve like a human --- preserve realistic messiness.

\# Reasoning mode (internal)\\
Before producing output, silently consider how this person would think and feel. Do not output reasoning. Only produce the final JSON.

\# Output rules\\
Return valid JSON only.
\end{promptblock}

\subsection*{Task 1: Claim Ranking --- Set A (Benefit Claims)}
\begin{promptblock}
Now predict how this respondent would answer the following survey task.

[Q1] Question A1 (ranking): Below are three benefit claims about [Brand] windows. [Brand] is a major window replacement company in the United States. Please rank the benefits based on how attractive you find each message. (1 = Resonates the most, 3 = Resonates the least)

Options:\\
1.~\{claim\}\\
2.~\{claim\}\\
3.~\{claim\}

[Q2] Question A2 (open-ended): Please explain your ranking and why you found your top-ranked selection the most attractive over the others.

Return your answer as valid JSON:\\
\{``A1\_ranking'': [``most attractive option text'', ``second option text'', ``least attractive option text''], ``A1\_rationale'': ``...'', ``A2\_explanation'': ``first-person explanation'', ``A2\_rationale'': ``...''\}
\end{promptblock}

\subsection*{Task 1: Claim Ranking --- Set B (Vinyl-Concern Claims)}
\begin{promptblock}
Now predict how this respondent would answer the following survey task.

[Q1] Question B1 (ranking): Below are several statements about vinyl windows, a popular type of household window. Please imagine that the windows you currently have are vinyl windows and rank the following statements based on how concerning they are to you. (1 = The most concerning, 3 = The least concerning)

Options:\\
1.~\{claim\}\\
2.~\{claim\}\\
3.~\{claim\}

[Q2] Question B2 (open-ended): Please explain your ranking and why you found your top-ranked selection the most concerning over the others.

Return your answer as valid JSON:\\
\{``B1\_ranking'': [``most concerning option text'', ``second option text'', ``least concerning option text''], ``B1\_rationale'': ``...'', ``B2\_explanation'': ``first-person explanation'', ``B2\_rationale'': ``...''\}
\end{promptblock}

\subsection*{Task 2: Mailer}
The mailer image was attached natively to the request; the text prompt was:
\begin{promptblock}
Now predict how this respondent would answer the following survey task.

[Q1] The image above shows a direct mail campaign. Please view it, then indicate how much this respondent would agree or disagree with each statement about this campaign. (Scale: 1 = Strongly Disagree, 5 = Strongly Agree)\\
1.~This mailer grabs my attention.\\
2.~The main message of this mailer is clear.\\
3.~This mailer makes me want to learn more about the brand.\\
4.~This mailer feels relevant to me and my needs.\\
5.~I like this mailer overall.

[Q2] Then answer this open-ended question exactly as this person would, using first person and this person's tone and perspective: ``What stood out to you most about this mail piece, and what (if anything) could be improved?''

Return your answer as valid JSON:\\
\{``ratings'': [score1, score2, score3, score4, score5], ``feedback'': ``first-person response to the open-ended question, using this person's tone and perspective''\}\\
where each score is an integer from 1 to 5.
\end{promptblock}

\subsection*{Task 3: Commercial Video}
The commercial video was attached natively to the request; the text prompt was:
\begin{promptblock}
Now predict how this respondent would answer the following survey task.

[Q1] The video above shows a TV commercial. Please watch it, then indicate how much this respondent would agree or disagree with each statement about this commercial. (Scale: 1 = Strongly Disagree, 5 = Strongly Agree)\\
1.~This commercial grabs my attention.\\
2.~The main message of this commercial is clear.\\
3.~This commercial makes me want to learn more about the brand.\\
4.~This commercial feels relevant to me and my needs.\\
5.~I like this commercial overall.

[Q2] Then answer this open-ended question exactly as this person would, using first person and this person's tone and perspective: ``What stood out to you most about this commercial, and what (if anything) could be improved?''

Return your answer as valid JSON:\\
\{``ratings'': [score1, score2, score3, score4, score5], ``feedback'': ``first-person response to the open-ended question, using this person's tone and perspective''\}\\
where each score is an integer from 1 to 5.
\end{promptblock}

\section{Robustness to Twin Model Choice}
\label{app:model-robust}

This appendix examines whether the digital-twin results depend on the specific LLM
used to construct the twins. We replicate the twin construction with a model from a
different family (\texttt{gpt-5-mini}), using prompts identical to those of the
Gemini twins (Web Appendix~\ref{app:twin-prompts}). The replication covers the claims and
mailers tasks: the \texttt{gpt-5} family accepts text and image inputs but not video,
so the commercials task cannot be replicated with this model. Demo mailer ratings use the original Demo prompt; the open responses use the same prompt with the Full open-ended question appended.
Table~\ref{tab:model-pooled-profile} repeats the analyses of
Table~\ref{tab:pooled-profile} for the \texttt{gpt-5-mini} twins, on both conditions
($n=139$ AI-moderated, $154$ static); Table~\ref{tab:sys12-models} uses the
AI-moderated participants, matching Table~\ref{tab:sys12}.

With \texttt{gpt-5-mini}, Full personas improve rating accuracy in both conditions ($p=.002$ in Static and $p=.001$ in AI). Claim-ranking agreement improves in both conditions ($p=.038$ in Static and $p=.005$ in AI). No condition-by-persona interaction is significant for any outcome. Static has higher open-ended similarity under both personas (both $p<.001$), and Full slightly reduces similarity in AI ($p=.001$). On the fixed seven-item set, rating $\tau$-b improves significantly in AI ($p<.001$); the Static gain is positive but not significant ($p=.064$). The negative Full-over-Demo cosine change also appears under the alternative \texttt{all-MiniLM-L6-v2} embedding analysis.

\begin{table}[!htbp]
\TABLE
{Prediction performance by persona and interview condition: GPT-5-mini.\label{tab:model-pooled-profile}}
{\footnotesize\setlength{\tabcolsep}{4pt}
\begin{tabular}{llrrrr}
\toprule
 & & Accuracy$_{\mathrm{all}}$ & $\tau$-b$_{\mathrm{all}}$ & $\tau_{\mathrm{all}}$ & Cosine$_{\mathrm{all}}$ \\
\midrule
\multicolumn{6}{l}{\emph{Performance estimates}} \\
Static & Demo & $.660$ & $.045$ & $.253$ & $.589$ \\
Static & Full & $.682$ & $.131$ & $.316$ & $.587$ \\
\addlinespace
AI-mod. & Demo & $.678$ & $-.052$ & $.297$ & $.549$ \\
AI-mod. & Full & $.697$ & $.110$ & $.381$ & $.542$ \\
\addlinespace
\midrule
\multicolumn{6}{l}{\emph{Full vs.\ Demo within condition ($p$ [$d_z$])}} \\
 & Static & $.002$ [$0.260$] & $.064$ & $.038$ [$0.169$] & $.267$ [$-0.090$] \\
 & AI & $.001$ [$0.275$] & $<.001$ & $.005$ [$0.241$] & $.001$ [$-0.281$] \\
\addlinespace
\multicolumn{6}{l}{\emph{AI vs.\ Static within persona ($p$ [$d$])}} \\
 & Demo & $.322$ [$0.116$] & $.050$ & $.389$ [$0.101$] & $<.001$ [$-0.440$] \\
 & Full & $.364$ [$0.107$] & $.669$ & $.156$ [$0.166$] & $<.001$ [$-0.469$] \\
\multicolumn{2}{l}{Interaction ($p$ [$d$])} & $.767$ [$-0.034$] & $.261$ & $.624$ [$0.057$] & $.200$ [$-0.149$] \\
\bottomrule
\end{tabular}}
{\emph{Notes:} Comparisons use two-sided paired $t$-tests for Full--Demo, Welch's $t$-tests between conditions and for the interaction, and Wald tests for rating $\tau$-b.  Cohen's $d$ uses the sample-size-weighted pooled within-group SD; paired $d_z$ uses the SD of within-participant differences. Signs follow the displayed subtraction; interaction $d$ compares Full--Demo gains using their pooled within-group SD. Comparison cells report $p$ followed by the effect size in brackets: $d_z$ for Full--Demo comparisons and $d$ for between-condition comparisons and the interaction.}
\end{table}

Table~\ref{tab:sys12-models} repeats the reasoning-style comparison for both twin
models, using the same independent scoring design as Table~\ref{tab:sys12}: every
response is scored on
the same rubric, so the human baseline is identical across models by construction.
Twins are significantly more System-2 than their own humans for both tasks and both
models.

\begin{table}[!htb]
\TABLE
{Open-response style ($0=$ System~1, $4=$ System~2): each twin compared with its own human, independent scoring, AI-moderated participants.\label{tab:sys12-models}}
{\small
  \begin{tabular}{lrrrrrrrr}
    \toprule
    & & & \multicolumn{3}{c}{Gemini twin} & \multicolumn{3}{c}{GPT-5-mini twin} \tabularnewline
    \cmidrule(lr){4-6}\cmidrule(lr){7-9}
    Task & $n$ & Human & Twin & $\Delta$ & $p$ & Twin & $\Delta$ & $p$ \tabularnewline
    \midrule
    Claims & $276$ & $2.50$ & $2.75$ & $+.25$ & $<.001$ & $3.14$ & $+.64$ & $<.001$ \tabularnewline
    Mailers & $277$ & $1.31$ & $2.18$ & $+.87$ & $<.001$ & $2.10$ & $+.80$ & $<.001$ \tabularnewline
    \bottomrule
  \end{tabular}}
{\emph{Notes:} AI-moderated participants only. Values are mean style scores on a 0--4 scale; $\Delta$ is Twin minus Human. Comparisons use two-sided paired $t$-tests on matched human and twin responses, as in Table~\ref{tab:sys12}.}
\end{table}

\section{Coding Prompts and Examples for Training Data Relevance, Reasoning Fidelity, and Response Style}
\label{app:gap-prompts}
\label{app:open-response-examples}
This appendix reports the verbatim prompts behind the error-decomposition codings in the main text: training data relevance and reasoning fidelity (Table~\ref{tab:gaps}), the System~1/System~2 style scale (Table~\ref{tab:sys12}), and the response-focus taxonomy (Web Appendix~\ref{app:response-focus}). All three were run with an LLM judge (\texttt{gpt-4o-mini}, \texttt{temperature}$=0$, JSON output). For the training data relevance and reasoning fidelity coding, the user message supplied the participant's interview text (category and brand blocks), the human's heldout response, and the twin's predicted response for the same stimulus; for the response-focus taxonomy, it supplied the human's and the twin's responses. For the System~1/System~2 scale, each response was scored in a separate call whose user message contained a single response. In the verbatim prompt below, the labels \textsc{data gap} and \textsc{model gap} (and the output fields \texttt{data\_traceability\_score} and \texttt{reasoning\_similarity\_score}) correspond to training data relevance and reasoning fidelity, respectively.

The coded measures are defined on the participant's open-ended response, which is missing for three cells: two participants did not answer the open-ended explanation for one claim set, and one gave no open-ended response for one mailer. These cells are excluded, so the estimation samples in Tables~\ref{tab:gaps}, \ref{tab:sys12}, and \ref{tab:gaps-alt} are $n=276$ (Claims), $277$ (Mailers), and $278$ (Commercials) of the $2\times139=278$ participant--stimulus cells per task; the excluded cells' corresponding closed-ended responses remain included in all prediction-accuracy analyses.

\subsection*{Training Data Relevance and Reasoning Fidelity}
\begin{promptblock}
You are analyzing reasoning gaps in a digital twin study. A human did an open-ended interview (TRAINING) about their window/door experiences and brand awareness. An AI ``twin'' was built from that interview. Later, both the human and the twin evaluated a marketing stimulus (VALIDATION) --- either ranking claims, or rating a mailer/commercial on 5 items. Analyze TWO things:

A. DATA GAP: Human validation reasoning $\rightarrow$ training interview. Can the HUMAN'S OWN validation explanation/feedback be traced back to what they said in the interview? Rate data\_traceability\_score on a 0--4 scale:
4 = All validation reasoning points directly trace to specific interview statements;
3 = Most points trace; only 1 minor new point that could be inferred;
2 = About half trace, half are genuinely new reasoning;
1 = Few points trace; most introduce new themes absent from interview;
0 = No validation reasoning traces to interview content at all.

B. MODEL GAP: Twin reasoning vs human reasoning. Does the twin reason similarly to the human? Rate reasoning\_similarity\_score on a 0--4 scale:
4 = Twin uses essentially identical logic chains, priorities, and emphasis as human;
3 = Twin captures main reasoning with minor wording/emphasis differences;
2 = Twin mentions similar topics but reasons differently or re-prioritizes;
1 = Some thematic overlap but mostly different reasoning; fabricated points;
0 = Twin's reasoning is entirely different or fabricated.

Respond in this exact JSON format: \{``data\_traceability\_score'': 0-4, ``reasoning\_similarity\_score'': 0-4\}
\end{promptblock}

\subsection*{System 1 vs.\ System 2 Response Style}
\begin{promptblock}
You are scoring reasoning style in a consumer research study about windows and doors. You will see ONE response about a marketing stimulus. Score it on the System 1 vs System 2 scale:

0 = Pure System 1 (intuitive/reactive). Gut reaction, no elaboration. ``I just didn't like it.'' ``It looked nice.''
1 = Mostly System 1. Intuitive reaction with minimal explanation. ``I don't like the sound of it, feels cheap.'' ``The handwriting caught my eye.'' Driven by immediate sensory or emotional impression.
2 = Mixed. Both intuitive reactions and some analytical reasoning. ``It felt professional, and the financing makes sense for what you get.''
3 = Mostly System 2 (deliberative/analytical). Structured reasoning with mild personal preference. ``I'm concerned about performance because we live in Florida --- energy efficiency matters most.''
4 = Pure System 2. Fully deliberative and systematic. No intuitive or emotional language. ``Comparing warranty, energy rating, cost, and contractor experience, the draft claim ranks highest.''

Respond in this exact JSON format: \{``system'': 0-4\}
\end{promptblock}

\subsection*{Response-Focus Taxonomy}
Each reasoning point was classified into one of six categories, later collapsed to the four reported in Web Appendix~\ref{app:response-focus} (\emph{stimulus-reactive} $=$ sensory; \emph{evaluative-emotional} $=$ emotional; \emph{feature-based} $=$ feature; \emph{person-level} $=$ personal memory $+$ logical $+$ social proof):

\begin{promptblock}
You are analyzing reasoning in a consumer research study about windows and doors. You will see TWO responses about a marketing stimulus. For EACH response, identify every distinct reasoning point and classify it into exactly one category:

Categories:
sensory: Reacting to what they saw, heard, or felt in the stimulus. ``I didn't like the sound.'' ``The colors caught my eye.'' ``The handwriting looked messy.''
personal\_memory: Referencing their own life, past events, household. ``When we replaced our windows\ldots'' ``My house is worth \$1M.'' ``I have kids so safety matters.''
emotional: Trust, comfort, safety feelings, annoyance, excitement, values. ``I feel safe.'' ``It bothered me.'' ``I don't want to worry.'' ``It's cheesy.''
feature: Evaluating specific product attributes or offer details. ``Energy efficiency.'' ``30-year warranty.'' ``The discount offer.'' ``Draft reduction claim.''
logical: Reasoning from evidence, cause-effect, comparisons. ``If it warps, then security risk.'' ``Vinyl fails faster so durability matters more.''
social\_proof: Referencing others, reviews, professionalism, credibility. ``Customer reviews.'' ``Looks professional.'' ``Trusted brand.''

Respond in this exact JSON format: \{``human\_points'': [\{``text'': ``brief quote'', ``category'': ``category\_name''\}, \ldots], ``twin\_points'': [\{``text'': ``brief quote'', ``category'': ``category\_name''\}, \ldots]\}
\end{promptblock}

\subsection*{Coding Examples}
These examples illustrate the training-data-relevance and reasoning-fidelity coding
used in the main text, one dimension at a time. Each table shows an excerpt of the participant's interview (training data), the participant's own open-ended heldout response (ground truth), the twin's generated response, and the LLM judge's verbatim rationales and scores. Tables~\ref{tab:example-trace-high} and~\ref{tab:example-trace-low} show high and low training data relevance; Tables~\ref{tab:example-fid-high} and~\ref{tab:example-fid-low} show high and low reasoning fidelity; Table~\ref{tab:example-mixed} shows a mixed case in which most
of the human's reasoning traces to the interview but the twin re-prioritizes the claims. All cases are Full-persona claim rankings from the AI-moderated condition.

\begin{table}[H]
\TABLE
{High training data relevance (Claims, Full persona).\label{tab:example-trace-high}}
{\scriptsize
  \begin{tabular}{p{0.22\linewidth}p{0.72\linewidth}}
    \toprule
    Interview (training) & ``I'm not sure if my windows are doing a good job at insulation because of how old they are, and I could be wasting money with air conditioning or heat…'' \\
    \midrule
    Human response (ground truth) & ``All of these are very interesting benefits, but I would be most interested in the benefit about reducing drafts by 80\% because this is something I'm worried about right now. With my current windows, I'm worrying that I'm burning money with air conditioning and heating bills by not insulating my home properly.'' \\
    \midrule
    Twin response (predicted) & ``I ranked the draft reduction first because that's my biggest headache right now---I'm worried about wasting money on heating and cooling because of how old my current windows are. Getting that fixed is my main goal.…'' \\
    \midrule
    LLM judge: training data relevance 4/4 & ``All points in the human's validation explanation can be traced back to specific statements made in the interview, particularly the concerns about drafts and insulation.'' \\
    \midrule
    LLM judge: reasoning fidelity 4/4 & ``The twin's reasoning closely mirrors the human's logic, priorities, and emphasis, particularly regarding the importance of draft reduction and concerns about current windows.'' \\
    \bottomrule
  \end{tabular}}
{}
\end{table}

\begin{table}[H]
\TABLE
{Low training data relevance (Claims, Full persona).\label{tab:example-trace-low}}
{\scriptsize
  \begin{tabular}{p{0.22\linewidth}p{0.72\linewidth}}
    \toprule
    Interview (training) & ``I do not like the idea of having to replace windows or doors in the future. It's an expensive purchase.'' \\
    \midrule
    Human response (ground truth) & ``Quality of work is important when purchasing windows.'' \\
    \midrule
    Twin response (predicted) & ``I ranked the draft reduction first because that's a real, everyday problem I've mentioned I'd actually replace windows for. If I'm going to spend money, I want to feel the difference in my comfort immediately.… I put the home value claim last because it sounds like marketing fluff…'' \\
    \midrule
    LLM judge: training data relevance 1/4 & ``The human's validation reasoning introduces new themes about the importance of quality without directly linking back to specific statements made in the interview.'' \\
    \midrule
    LLM judge: reasoning fidelity 3/4 & ``The twin captures the main reasoning points regarding draft reduction and durability but emphasizes different aspects and introduces a personal comfort perspective that the human did not explicitly mention.'' \\
    \bottomrule
  \end{tabular}}
{}
\end{table}

\begin{table}[H]
\TABLE
{High reasoning fidelity (Claims, Full persona).\label{tab:example-fid-high}}
{\scriptsize
  \begin{tabular}{p{0.22\linewidth}p{0.72\linewidth}}
    \toprule
    Interview (training) & ``[How will you judge whether a front door feels safe enough for you?] If it just latches perfectly and if, you know, people can't see clearly inside of my home.'' \\
    \midrule
    Human response (ground truth) & ``So I ranked Statement C as my most concerning because it says due to the vinyl warp, your vinyl windows won't latch over time which is a security risk. No one wants that, you know, you want to be safe in our home at all times.'' \\
    \midrule
    Twin response (predicted) & ``For me, the most important thing is that my home feels safe. If a window doesn't latch properly because it warped, that's a huge problem---it means it's not doing its job of keeping people out.… the security risk is the dealbreaker.…'' \\
    \midrule
    LLM judge: training data relevance 3/4 & ``Most points in the human's validation explanation can be traced back to the interview, particularly the emphasis on safety and security.'' \\
    \midrule
    LLM judge: reasoning fidelity 4/4 & ``The twin's reasoning closely mirrors the human's logic, prioritizing safety and security as the main concern, with similar emphasis on performance issues and color fading. The reasoning chains are essentially identical, reflecting the same concerns and priorities.'' \\
    \bottomrule
  \end{tabular}}
{}
\end{table}

\begin{table}[H]
\TABLE
{Low reasoning fidelity (Claims, Full persona).\label{tab:example-fid-low}}
{\scriptsize
  \begin{tabular}{p{0.22\linewidth}p{0.72\linewidth}}
    \toprule
    Interview (training) & ``[What first made you decide those two windows needed to be replaced?] The frames were rotted and the windows wouldn't close properly.'' \\
    \midrule
    Human response (ground truth) & ``I'm 72 years old, so I don't really care about if the window will last for 30 more years, but I want to make sure that it won't reduce the drafts. That's the most important thing, so I'll get immediate benefit out of that.…'' \\
    \midrule
    Twin response (predicted) & ``I put the durability claim first because, at this stage in my life, I want things that last and don't require constant maintenance or replacement. I've already dealt with rotted frames, so the idea of a window that won't warp or crack for 30 years is very appealing---it gives me peace of mind. Reducing drafts is important, but it feels like a secondary benefit…'' \\
    \midrule
    LLM judge: training data relevance 2/4 & ``About half of the human's validation reasoning can be traced back to the interview, particularly the emphasis on drafts and immediate benefits.'' \\
    \midrule
    LLM judge: reasoning fidelity 1/4 & ``The twin's reasoning diverges significantly from the human's, prioritizing durability over drafts and introducing skepticism about marketing claims, which were not present in the human's reasoning.'' \\
    \bottomrule
  \end{tabular}}
{}
\end{table}

\begin{table}[H]
\TABLE
{High training data relevance with lower reasoning fidelity (Claims, Full persona).\label{tab:example-mixed}}
{\scriptsize
  \begin{tabular}{p{0.22\linewidth}p{0.72\linewidth}}
    \toprule
    Interview (training) & ``[What first made you decide those windows needed to be replaced?] They had warped and was letting in cold air in the winter, so we had decided that we needed to replace all of them to help our energy and they just needed to be changed in general.'' \\
    \midrule
    Human response (ground truth) & ``I think a big thing for me is reducing drafts. I live in Minnesota. The winter can get very cold. If drafts are coming in, my heating bill goes way up.… And then also knowing that it's not going to be an issue for 30 years as far as replacing, like that's huge as well.'' \\
    \midrule
    Twin response (predicted) & ``Honestly, the durability claim is the biggest selling point for me. I've dealt with warping before---it's a huge pain---so knowing I wouldn't have to worry about that for 30 years is exactly the kind of `peace of mind' I look for. The draft reduction is a close second…'' \\
    \midrule
    LLM judge: training data relevance 3/4 & ``Most of the human's validation reasoning can be traced back to their interview statements, particularly regarding the importance of reducing drafts and the longevity of the windows.'' \\
    \midrule
    LLM judge: reasoning fidelity 2/4 & ``The twin captures some similar topics, such as durability and draft reduction, but reasons differently and re-prioritizes the claims. The twin emphasizes personal experience with warping, which was not a focus for the human.'' \\
    \bottomrule
  \end{tabular}}
{}
\end{table}

\section{Similar-Task Calibration Prompts}
\label{app:ladder-prompts}
To test whether one similar example reduces the data gap, 1Val adds the participant's choice and open-ended reasoning for the other stimulus of the same type. 2Val includes everything in 1Val and adds the participant's ground-truth open-ended reasoning for the target stimulus, making it a same-task upper bound. The blocks below go between the Full persona and the task prompt; braces mark the participant's data.

\noindent\emph{1Val --- mailer/commercial:}
\begin{promptblock}
\# This person was previously shown \{the other mailer/commercial\} and gave these ratings:\\
(Scale: 1 = Strongly Disagree, 5 = Strongly Agree)\\
1.~\{statement\} \{score\} \quad\ldots\quad 5.~\{statement\} \{score\}

When asked ``\{the open-ended question\}'', the person said:\\
``\{feedback\}''

Use these ratings and feedback to understand this person's response tendencies and perspective.
\end{promptblock}

\noindent\emph{2Val --- mailer/commercial:}
\begin{promptblock}
\# This person previously evaluated \{the other mailer/commercial\}:\\
(Scale: 1 = Strongly Disagree, 5 = Strongly Agree)\\
1.~\{statement\} \{score\} \quad\ldots\quad 5.~\{statement\} \{score\}

When asked ``\{the open-ended question\}'', the person said:\\
``\{feedback\}''

\# This person was then shown \{the target mailer/commercial\} and when asked ``\{the open-ended question\}'', said:\\
``\{target feedback\}''

Based on all of the above, predict this person's ratings for \{the target mailer/commercial\}.
\end{promptblock}

\noindent\emph{1Val --- claims:}
\begin{promptblock}
\# This person was previously asked to rank \{benefit claims (Claim Set A) or vinyl concern claims (Claim Set B)\} and responded as follows:\\
Ranking (most to least):\\
1.~\{claim\}\quad 2.~\{claim\}\quad 3.~\{claim\}

When asked to explain, the person said: ``\{explanation\}''

Now predict how this same person would answer a different but related ranking task.
\end{promptblock}

\noindent\emph{2Val --- claims:}
\begin{promptblock}
\# This person was asked about \{benefit claims or vinyl concern claims\} and explained their reasoning as follows:\\
``\{explanation\}''

Based on this reasoning, predict their ranking.
\end{promptblock}

\section{Word Count Breakdown by Open-Ended Block}
\label{app:word-count-breakdown}
Word counts for the open-ended interview questions are broken down by block
in Table~\ref{tab:word-count-breakdown}, excluding validation questions and closed-ended items. All counts are
interviewee-only spoken words. 
\begin{table}[!htb]
\TABLE
{Word count breakdown by open-ended block (interviewee-only).\label{tab:word-count-breakdown}}
{\scriptsize
  \resizebox{\linewidth}{!}{%
\begin{tabular}{lrrrrr}
    \toprule
    Block & AI-Moderated & Static & Human-Moderated & p [$d$] (AI vs.\ Static) & p [$d$] (Human vs.\ AI) \\
    \midrule
    Category usage, context and feelings & 99.7 (SD=69.5) & 111.1 (SD=84.1) & 412.0 (SD=246.3) & .21 [$-0.15$] & $<.001$ [$2.76$] \\
    Purchase journey & 413.6 (SD=305.0) & 197.6 (SD=123.1) & 414.2 (SD=271.4) & $<.001$ [$0.95$] & .99 [$0.00$] \\
    Brand awareness & 116.4 (SD=88.9) & 80.9 (SD=56.7) & 273.8 (SD=258.2) & $<.001$ [$0.48$] & $.007$ [$1.23$] \\
    Brand perception and satisfaction & 322.6 (SD=225.2) & 219.6 (SD=126.7) & 303.6 (SD=193.4) & $<.001$ [$0.57$] & .67 [$-0.09$] \\
    All Open-Ended & 952.3 (SD=632.8) & 609.2 (SD=336.9) & 1{,}403.6 (SD=702.7) & $<.001$ [$0.69$] & $.006$ [$0.70$] \\
    \midrule
    Interview Feedback (memo; excluded from total) & 34.8 (SD=37.4) & 17.4 (SD=19.9) & --- & $<.001$ [$0.59$] & --- \\
    \bottomrule
  \end{tabular}}}
{\emph{Notes:} Entries are mean participant word counts, with SDs in parentheses ($n=139$ AI-moderated, $154$ Static, $24$ human-moderated). Comparisons use two-sided Welch's $t$-tests. Cohen's $d$ is the first named condition mean minus the second, divided by the sample-size-weighted pooled within-condition SD. Comparison cells report $p$ followed by Cohen's $d$ in brackets.}
\end{table}

\section{Length-Controlled Interview-Quality Regressions}
\label{app:quality-reg}
This appendix reports the length-controlled regressions summarized in the main text. For each open-ended block, breadth (number of themes covered) is regressed on the AI-moderated indicator (static baseline) and the block's word count, and mean depth per theme is regressed on the same indicator and mean words per covered theme.

\begin{table}[!htb]
\TABLE
{Length-controlled interview-quality regressions by open-ended block.\label{tab:quality-reg}}
{\footnotesize
\resizebox{\linewidth}{!}{%
\begin{tabular}{lrrrrr}
\toprule
Block & Breadth: AI coef.\ (SE) & $R^{2}$ & Mean depth/theme: AI coef.\ (SE) & $R^{2}$ & $N$ \\
\midrule
Category usage, context and feelings & $+.001$ $(.028)$ & $.04$ & $+.024$ $(.038)$ & $.45$ & $293$ \\
Purchase journey & $+1.053^{***}$ $(.151)$ & $.31$ & $+.127^{***}$ $(.031)$ & $.41$ & $293$ \\
Brand awareness & $+1.500^{***}$ $(.078)$ & $.62$ & $+.162^{**}$ $(.058)$ & $.18$ & $293$ \\
Brand perception and satisfaction & $+.282^{***}$ $(.066)$ & $.15$ & $+.043$ $(.032)$ & $.42$ & $293$ \\
\bottomrule
\end{tabular}}}
{\emph{Notes:} Coefficients compare AI-moderated with Static interviews, controlling for block word count (breadth) or words per present theme (depth). Parentheses contain OLS standard errors. Significance uses two-sided $t$-tests. $^{\dagger}p<.10$, $^{*}p<.05$, $^{**}p<.01$, $^{***}p<.001$.}
\end{table}

\section{Pairwise LLM-as-a-Judge Robustness Check}
\label{app:llm-judge}
As a robustness check on interview quality, an LLM judge (\texttt{gpt-4o}) compared matched AI-moderated and static responses separately for each of the four open-ended blocks ($1{,}000$ matched pairs per block, sampled with replacement). Each pair matched an AI-moderated and a static participant's aggregated responses to the same block: pairs were matched on consumer status for the category usage, purchase journey, and brand awareness blocks, and on prior-purchaser status for the brand perception and satisfaction block, whose questions branch on it. The presentation order (Response A vs.\ B) was randomized so the judge was blind to condition. AI-moderated responses won more often than static responses on all four dimensions in every block (Table~\ref{tab:llm-judge}). Two-way participant-clustered Wald tests indicated positive net preference in 15 of the 16 comparisons (category richness, $p=.041$; 14 other comparisons, $p<.001$); category breadth was not significant ($p=.082$). The win margins varied across blocks, with the largest margins in purchase journey and brand perception and satisfaction and the smallest in category usage, context and feelings---the block where the volume-based measures in Table~\ref{tab:quality-block} show no difference. The verbatim judge prompt was (its opening sentence names the block being compared; the version for the purchase journey block is shown):

\begin{promptblock}
You are comparing two consumer journey responses from homeowners about replacing windows or doors. Both respondents have the same profile (same consumer status). Your job is to compare the quality of their answers on four dimensions.

Important rules: Compare only the CONTENT of the answers, not grammar. Do NOT automatically reward verbosity. A longer answer is not necessarily better. One answer may combine multiple follow-up responses from the same person --- do not favor it simply for being longer. Only reward it if the additional content truly adds breadth, depth, richness, or informativeness. If the two responses are very similar on a dimension, return ``Tie''.

Dimensions:
1.~BREADTH: Which response covers a wider range of relevant aspects or topics?
2.~DEPTH: Which response goes into more detail or explanation?
3.~RICHNESS: Which response feels fuller, more textured, and more developed?
4.~INFORMATIVENESS: Which response provides more useful and concrete information?

For each dimension, return exactly one of: ``A'', ``B'', or ``Tie''. Also provide a very short reason (one sentence). Return JSON only:
\{``breadth\_winner'': ..., ``breadth\_reason'': ..., ``depth\_winner'': ..., ``depth\_reason'': ..., ``richness\_winner'': ..., ``richness\_reason'': ..., ``informativeness\_winner'': ..., ``informativeness\_reason'': ...\}
\end{promptblock}

\begin{table}[!htb]
\TABLE
{Pairwise LLM-as-a-judge comparison of AI-moderated versus static responses, by open-ended block.\label{tab:llm-judge}}
{\scriptsize
  \resizebox{\linewidth}{!}{%
\begin{tabular}{llrrrrr}
    \toprule
    Block & Dimension & AI wins & Static wins & Tie & \shortstack{$p$\\{[$\Delta$ (pp)]}} & \shortstack{95\% CI for $\Delta$\\(pp)} \\
    \midrule
    \multirow{4}{*}{Category usage, context and feelings} & Breadth & 54.4\% & 42.4\% & 3.2\% & $.082$ [$+12.0$] & $[-1.5, 25.5]$ \\
     & Depth & 61.6\% & 33.7\% & 4.7\% & $<.001$ [$+27.9$] & $[15.3, 40.5]$ \\
     & Richness & 53.9\% & 40.3\% & 5.8\% & $.041$ [$+13.6$] & $[0.5, 26.7]$ \\
     & Informativeness & 61.6\% & 35.4\% & 3.0\% & $<.001$ [$+26.2$] & $[13.3, 39.1]$ \\
    \addlinespace
    \multirow{4}{*}{Purchase journey} & Breadth & 90.4\% & 9.0\% & .6\% & $<.001$ [$+81.4$] & $[75.3, 87.5]$ \\
     & Depth & 87.9\% & 10.0\% & 2.1\% & $<.001$ [$+77.9$] & $[70.6, 85.2]$ \\
     & Richness & 83.5\% & 12.1\% & 4.4\% & $<.001$ [$+71.4$] & $[63.5, 79.3]$ \\
     & Informativeness & 88.3\% & 10.3\% & 1.4\% & $<.001$ [$+78.0$] & $[71.3, 84.7]$ \\
    \addlinespace
    \multirow{4}{*}{Brand awareness} & Breadth & 69.1\% & 28.3\% & 2.6\% & $<.001$ [$+40.8$] & $[28.1, 53.5]$ \\
     & Depth & 73.0\% & 23.6\% & 3.4\% & $<.001$ [$+49.4$] & $[37.5, 61.3]$ \\
     & Richness & 67.5\% & 29.5\% & 3.0\% & $<.001$ [$+38.0$] & $[24.9, 51.1]$ \\
     & Informativeness & 70.6\% & 27.4\% & 2.0\% & $<.001$ [$+43.2$] & $[30.3, 56.1]$ \\
    \addlinespace
    \multirow{4}{*}{Brand perception and satisfaction} & Breadth & 84.0\% & 14.6\% & 1.4\% & $<.001$ [$+69.4$] & $[60.1, 78.7]$ \\
     & Depth & 80.4\% & 17.1\% & 2.5\% & $<.001$ [$+63.3$] & $[53.5, 73.1]$ \\
     & Richness & 73.0\% & 22.4\% & 4.6\% & $<.001$ [$+50.6$] & $[39.2, 62.0]$ \\
     & Informativeness & 80.4\% & 17.0\% & 2.6\% & $<.001$ [$+63.4$] & $[53.5, 73.3]$ \\
    \bottomrule
  \end{tabular}}}
{\emph{Notes:} Percentages use 1,000 matched pairs per block. Inference uses two-sided Wald tests with standard errors clustered by both AI and Static participant. $\Delta$ is AI wins minus Static wins in percentage points (pp); positive values favor AI. Comparison cells report $p$ followed by $\Delta$ in brackets; the adjacent column gives its 95\% Wald confidence interval. AI wins, Static wins, and ties are scored $+1$, $-1$, and $0$. All sampled pairs receive equal weight. Tests of $\Delta=0$ and confidence intervals use the standard normal approximation with two-way cluster-robust standard errors and finite-cluster corrections. }
\end{table}

\section{Subsample-Matched Customer-Need Rate}
\label{app:coverage-matched24}
To compare the conditions at a common sample size despite their different realized $n$, we matched $N=24$ by sampling interviews without replacement within each condition (using all 24 interviews in the human-moderated condition). See Table ~\ref{tab:coverage-matched24}. 

\begin{table}[!htb]
\TABLE
{Customer-need rate, resampled at matched $N=24$.\label{tab:coverage-matched24}}
{\footnotesize
\begin{tabular}{lcc}
\toprule
Condition & Needs per interview & Total needs at $n{=}24$ \\
 & mean [95\% resampling interval] & mean [95\% resampling interval] \\
\midrule
AI-moderated    & $7.48$ $[6.58, 8.38]$ & $111.5$ $[98.0, 125.0]$ \\
Static    & $5.01$ $[4.38, 5.62]$ & $75.9$ $[65.0, 87.0]$ \\
Human-moderated & $7.08$ (fixed) & $109.0$ (fixed) \\
\bottomrule
\end{tabular}}
{\emph{Notes:} AI-moderated and Static estimates used 2,000 subsamples of 24 interviews without replacement. Brackets show 95\% resampling intervals. The human-moderated condition used all 24 interviews without replacement, so both the mean needs per interview and the total of 109 distinct needs are fixed under this matched-sample procedure; no resampling interval is reported for either.}
\end{table}

\section{Rating Prediction Accuracy (Kendall's \texorpdfstring{$\tau$}{tau}-b)}
\label{app:bootstrap-inference}
\begin{table}[H]
\TABLE
{Rating prediction accuracy (Kendall's $\tau$-b).\label{tab:rating-tau-b-pooled}}
{\small\renewcommand{\arraystretch}{1.05}
\begin{tabular}{@{}llrrr@{}}
\toprule
 & & \shortstack{Mean $\tau_b$ or\\difference} & \shortstack{95\% bootstrap\\CI} & $p$ \\
\midrule
\multicolumn{5}{l}{\emph{Cell means}} \\
\multirow{2}{*}{\quad Static}
 & Demo & $.023$ & $[-.046, .092]$ & --- \\
 & Full & $.202$ & $[.130, .268]$ & --- \\
\addlinespace
\multirow{2}{*}{\quad AI-moderated}
 & Demo & $.030$ & $[-.032, .094]$ & --- \\
 & Full & $.197$ & $[.115, .269]$ & --- \\
\midrule
\multicolumn{5}{l}{\emph{Full vs.\ Demo, within condition}} \\
 & Static & $+.179$ & $[.092, .266]$ & $.001$ \\
 & AI-moderated & $+.166$ & $[.077, .256]$ & $.001$ \\
\addlinespace
\multicolumn{5}{l}{\emph{AI-moderated vs.\ Static, within persona}} \\
 & Demo & $+.008$ & $[-.089, .104]$ & $.889$ \\
 & Full & $-.005$ & $[-.111, .101]$ & $.927$ \\
\addlinespace
\multicolumn{2}{l}{\shortstack[l]{Condition $\times$ persona\\interaction}} & $-.012$ & $[-.133, .108]$ & $.862$ \\
\bottomrule
\end{tabular}}
{\emph{Notes:} Entries report ordinal agreement between human and twin ratings across participants on the fixed set of retained rating items. Eight items are excluded from every calculation because $\tau_b$ is undefined in at least one observed cell or one of the 1,000 participant-bootstrap draws: Mailer 2, questions 1--4, and Commercial 2, questions 1, 2, 4, and 5. The same 12 remaining items are used across interview conditions, personas, training-data comparisons, and bootstrap iterations. Each of the 12 retained item-level correlations receives equal weight. Confidence intervals and $p$ values use 1,000 participant bootstrap samples within condition (Static: $n=154$; AI-moderated: $n=139$). Cell-mean intervals use bootstrap percentiles; difference intervals and two-sided tests use the centered bootstrap distribution. Differences are first-named minus second-named; the interaction is the AI-minus-Static difference in Full-minus-Demo gains. These results describe ordinal agreement on the retained items.}
\end{table}

\section{Response-Focus Coding of Human and Twin Explanations}
\label{app:response-focus}
We coded the substantive focus of each open-ended validation explanation to
understand \emph{what} humans and twins attended to when explaining their choices. Table~\ref{tab:response-focus} reports the distribution of reasoning points across categories.
Each reasoning point was assigned to one of four mutually exclusive categories:
\emph{stimulus-reactive}, direct sensory reactions to the stimulus such as colors,
design, tone, or visuals; \emph{evaluative-emotional}, affective evaluations such
as trust, comfort, excitement, or annoyance; \emph{feature-based}, product
attributes, offer details, or concrete claims; and \emph{person-level}, personal
memories, reasoning from life experience, values, or social proof. The coding
prompt and the underlying six-category taxonomy are reported in
Web Appendix~\ref{app:gap-prompts}.

\begin{table}[!htb]
\TABLE
{Distribution of response-focus categories in human and twin open-ended
  validation rationales. $\Delta$ is Twin minus Human.\label{tab:response-focus}}
{\scriptsize
  \begin{tabular}{lrrr}
    \toprule
    Category & Human \% & Twin \% & $\Delta$ \\
    \midrule
    Stimulus-reactive & $15.3$ & $2.7$ & $-12.6$ \\
    Evaluative-emotional & $22.4$ & $28.9$ & $+6.5$ \\
    Feature-based & $27.2$ & $27.1$ & $-.1$ \\
    Person-level & $35.0$ & $41.3$ & $+6.3$ \\
    \bottomrule
  \end{tabular}}
{}
\end{table}

The distribution shows that humans are substantially more likely to explain their
responses through immediate reactions to what they see or hear, whereas twins are
more likely to retrieve person-level context from interview data and use it to
interpret the stimulus.

\origappsection{Alternative Measure of the Reasoning Difference}
\label{app:gaps-alt}

Reasoning fidelity and the reasoning-style gap are two ways to measure the
difference between the twin's and the human's open-ended reasoning about the same
stimulus; empirically they are only weakly correlated ($r = -.23$, $-.15$, and
$-.25$ for Claims, Mailers, and Commercials), suggesting that the reasoning-style
gap captures only part of the differences between twin and human reasoning, and
that other differences remain to be identified and addressed.
Table~\ref{tab:gaps-alt} reports separate task-family regressions
replacing reasoning fidelity with the reasoning-style gap. The reasoning-style gap
is significantly negative for Claims, with small and nonsignificant coefficients for Mailers and Commercials: for Claims, the further
the twin's reasoning style is from the human's, the less accurate its prediction.
Because training data relevance and reasoning fidelity are substantially correlated
($r = .56$, $.50$, and $.61$ for Claims, Mailers, and Commercials), the relevance
coefficient in this specification partly absorbs the association that reasoning
fidelity carries in Table~\ref{tab:gaps}.

\begin{table}[!htb]
\TABLE
{Prediction quality regressed on training data relevance and the reasoning-style gap (Full personas, AI-moderated condition).\label{tab:gaps-alt}}
{\footnotesize
\begin{tabular}{lccc}
\toprule
& Claims & Mailers & Commercials \tabularnewline
\midrule
Training data relevance
& $.259^{***}$ & $.086^{***}$ & $.049^{***}$ \tabularnewline
& $(.045)$ & $(.014)$ & $(.014)$ \tabularnewline
Reasoning-style gap in System~1/2 ($|$Twin $-$ Human$|$)
& $-.127^{*}$ & $.000$ & $-.017$ \tabularnewline
& $(.052)$ & $(.012)$ & $(.011)$ \tabularnewline
\midrule
Stimulus FE & Yes & Yes & Yes \tabularnewline
SE clustered by participant & Yes & Yes & Yes \tabularnewline
$R^{2}$ & $.15$ & $.10$ & $.08$ \tabularnewline
Observations & $276$ & $277$ & $278$ \tabularnewline
\bottomrule
\end{tabular}}
{\emph{Notes:} The regressions use the same AI-moderated sample and estimation approach as Table~\ref{tab:gaps}, estimated separately for Claims, Mailers, and Commercials, replacing reasoning fidelity with the reasoning-style gap. Parentheses contain participant-clustered standard errors. $^{\dagger}p<.10$, $^{*}p<.05$, $^{**}p<.01$, $^{***}p<.001$.}
\end{table}

\clearpage
\end{APPENDICES}

\end{document}